\documentclass[a4paper,10pt,twoside]{article}
\usepackage{amsmath}
\usepackage{amsthm}
\usepackage{amssymb}
\usepackage{bm}
\usepackage{dsfont}
\usepackage[plainpages=false,pdfpagelabels]{hyperref}
\usepackage{pgf,tikz}
\usetikzlibrary{arrows}
\usepackage[numbers]{natbib}

\theoremstyle{plain}
\newtheorem{theorem}{Theorem}[section]
\newtheorem{proposition}[theorem]{Proposition}
\newtheorem{lemma}[theorem]{Lemma}

\theoremstyle{definition}
\newtheorem{definition}[theorem]{Definition}

\newtheorem{remark}[theorem]{Remark}

\newcommand{\K}{\mathbb{K}}
\newcommand{\J}{\mathbb{J}}
\newcommand{\R}{\mathbb{R}}

\renewcommand{\H}{\mathbb{H}}
\renewcommand{\S}{\mathbb{S}}
\renewcommand{\O}{\mathbb{O}}
\renewcommand{\P}{\mathbb{P}}

\newcommand{\FF}{\mathbb{F}}

\newcommand{\EP}{ {\mathbb E}_{\mathbb{P}}}

\newcommand{\Aset}{\mathcal{A}}

\newcommand{\Filt}{\mathcal{F}}

\newcommand{\Mset}{\mathcal{M}}
\newcommand{\Eset}{\mathcal{E}}
\newcommand{\Strat}{\mathcal{S}}
\renewcommand{\Game}{\mathcal{G}}

\newcommand{\norm}[1]{\left\|#1\right\|}

\newcommand{\proj}[2]{\pi_{#2}\left(#1\right)}

\newcommand\I{\mathds{1}}

\title{Unilaterally Competitive Multi-Player Stopping Games}

\author{Ivan Guo
\footnote{School of Mathematics and Statistics, University of Sydney, NSW 2006, Australia --- \texttt{ivan.guo@sydney.edu.au}}
\footnote{This work was supported by Australian Research Council's Discovery Projects funding scheme (DP120100895).}
}

\date{\today}

\begin{document}
\maketitle

\begin{abstract}
A multi-player competitive Dynkin stopping game is constructed, extending the work of Guo and Rutkowski \cite{guo2012zero}. Each player can either exit the game for a fixed payoff, determined a priori, or stay and receive an adjusted payoff depending on the decision of other players. The single period case is shown to be ``weakly unilaterally competitive'' (Kats and Thisse \cite{kats1992unilaterally} and De Wolf \cite{de1999optimal}). We present an explicit construction of the unique value at which Nash and optimal equilibria are attained. Multiple period generalisations are explored. The game has interpretations in economic and financial contexts, for example, as a consumption model with bounded resources. It also serves as a starting point to the construction of multi-person financial game options. In particular, the concept of optimal equilibria becomes pivotal in the pricing of the game options via super-replication.

\medskip
\noindent \textbf{Keywords:} Dynkin game, stopping game, $n$-player game, optimal equilibrium, game option.\\
\noindent \textbf{AMS Subject Classification:} 91A06, 91A10, 91A15, 91A50, 60G40.
\end{abstract}

\section{Introduction}

A Dynkin stopping game, first introduced by Dynkin \cite{dynkin1969game}, is a zero-sum, optimal stopping game between two players where each player can stop the game for a payoff observable at that time. Much research has been done on this as well as its related problems, e.g., \cite{cvitanic1996backward,ekstrom2008optimal,hamadene2010continuous,laraki2010equilibrium,peskir2009optimal,rosenberg2001stopping,solan2001quitting,solan2003deterministic,touzi2002continuous}. One application of Dynkin games is in game contingent claims, or game option, as defined by Kifer \cite{kifer2000game}, who proved the existence and uniqueness of its value. Further works, such as Hamad\`{e}ne and Zhang \cite{hamadene2010continuous}
and Kallsen and K\"uhn \cite{kallsen2004pricing}, studied various techniques in its pricing.

Various formulations of multi-player Dynkin games exist in literature. For example, Solan and Vieille \cite{solan2001quitting} introduced a quitting game which terminates when any player chooses to quit, then each player receives a payoff depending the set of players quitting the game. Under certain payoff conditions, a subgame perfect uniform $\epsilon$-equilibrium using cyclic strategies can be found. In Solan and Vieille \cite{solan2003deterministic}, another version is presented, in which the players are given the opportunity to stop the game in a turn-based fashion. A subgame perfect $\epsilon$-equilibrium was again shown to exist and consisted of pure strategies when the game is not degenerate. Hamad\`{e}ne and Hassani \cite{hamadene2012multi} presented a non-zero sum stopping game where each player has his own separate payoff processes. These processes are independent of the other players' decisions, so in the event where a player does not stop first, his payoff does not depend the exact set of players who stopped.

We aim to generalise the Dynkin stopping game to more than two players in a natural way which allows for the construction of a multi-person financial game option. Guo and Rutkowski \cite{guo2012zero} introduced a zero-sum, simultaneous Dynkin game, with a focus on designing the \emph{dependencies} between the payoffs of all players and their stopping decisions. In effect, it is modelling a multilateral ``contract'' where all the players are competing for a fixed total sum of wealth. Each player can either exit or terminate the contract for a predetermined benefit, or do nothing and receive an adjusted benefit, reflecting the discrepancies caused by any exiting decisions. These adjustments ensure that the total wealth is fixed.

This paper extends the results of Guo and Rutkowski \cite{guo2012zero} to games which are not necessarily zero-sum, but still retains the ``weakly unilaterally competitive'' (or WUC) property introduced by Kats and Thisse \cite{kats1992unilaterally} and De Wolf \cite{de1999optimal}. As discussed Pruzhansky \cite{pruzhansky2011some}, Nash equilibria are not always adequate as a solution concept. This certainly occurs in the valuation of the game options, as the Nash equilibria payoffs cannot be guaranteed. Instead, we formally introduce the stronger ``optimal equilibria'', or Nash equilibria with maximin strategies, which induces a unique value for the game. The WUC property ensures that all Nash equilibria are also optimal equilibria.

The main results of the paper are Theorems \ref{thmsoln} and \ref{thmsoln3}, which proves the existence of the value by explicit construction, and expresses it as the projection onto a simplex under an appropriate choice of inner product. The construction also produces a pure strategy optimal equilibrium.

Several extensions are discussed. All single period results can be immediately applied to the stochastic case where both terminal and exercise payoffs are random, as long as expectations are incorporated into the definitions of solution and equilibria. Also, two multiple period generalisations are studied, including a recursive stopping game and a quitting game. The recursive stopping game can be readily applied to multi-person financial game options, where the properties of the optimal equilibrium become imperative in the pricing arguments. Details will be presented in an upcoming paper. The quitting game is a variant which cannot be stopped early, but each player can choose to quit at any time. Optimal equilibria are constructed for deterministic case and subgame perfect optimal equilibria are constructed if the game is perfect information.

Apart from multi-person financial game options, the game presented here may be interpreted in other economic and financial contexts, for example, as a consumption model with bounded resources. It serves as a starting point to a particular class of competitive multi-player games. Many more multiple period and continuous time generalisations are possible, and they are under further research.

Section \ref{secoptiequi} of the paper establishes some preliminary results in game theory and introduces the optimal equilibrium. 
Section \ref{secdeterm} constructs the single period game and proves the existence and uniqueness of the value.
Section \ref{secprojection} revisits the results of Guo and Rutkowski \cite{guo2012zero} and applies them to non-zero sum settings. The value is constructed using projection.
Section \ref{secmultiperiod} applies the results to the stochastic case, and then examines two multiple period possibilities.

\section{Optimal Equilibrium}\label{secoptiequi}
This section will discuss several game theory concepts, while keeping track of two main focuses: To find solution concepts applicable to the pricing of \emph{financial game options}, as well as conditions characterising the idea of \emph{competitiveness} in multi-person games.

Consider a game $\Game$ with $m$ \emph{players},  enumerated by the indices $1,2,\ldots,m$. The set of all players is denoted by $\Mset$. Each player $k$ can choose a \emph{strategy} $s_k\in\Strat_k$, and the $m$-tuples of strategies $s=\big[s_1,\ldots,s_m\big]\in\Strat$ are \emph{strategy profiles}. Given a strategy profile $s$, it is possible to compute a vector of \emph{payoff} functions $\bm{V}(s)=[V_1(s),\ldots,V_m(s)]$ for the players, with larger payoffs being more desirable.

As introduced by Nash \cite{nash1951non}, a strategy profile $s^*\in\Strat$ is referred to as a \emph{Nash equilibrium}, or simply an \emph{equilibrium}, if no single player can improve his payoff by altering his own strategy. Formally, for each $k\in\Mset$,
\[
V_k\left(\big[s_k^*, s_{-k}^*\big]\right)\geq V_k\left(\big[s_k, s_{-k}^*\big]\right), \quad \forall \, s_k\in\Strat_k.
\]
A Nash equilibrium represents a state which no player would deviate from. It gives some intuition to the value of the game, but in the context of game options, there are several deficiencies. The valuation of financial options involves replication of payoffs. But in general, this is not possible without knowing the action of the other players. One cannot assume that the other players will converge towards Nash equilibria, so the equilibrium payoffs are not guaranteed. Furthermore a game may have several equilibria leading to different payoff values, and it's not always clear which one should be chosen.

We formally introduce a new, stronger concept to address these issues.
\begin{definition}[Optimal Equilibrium] \label{optiequilibrium}
A strategy profile $s^*\in\Strat$ is called an \emph{optimal equilibrium} if, for each $k\in\Mset$,
\begin{gather*}
V_k\left(\big[s_k^*, s_{-k}\big]\right)\geq V_k\left(\big[s_k^*, s_{-k}^*\big]\right)
\geq V_k\left(\big[s_k, s_{-k}^*\big]\right), \quad \forall \, s_k\in\Strat_k,\, \forall \, s_{-k}\in \Strat_{-k}.
\end{gather*}
\end{definition}

An optimal equilibrium is essentially a saddle point. It has the properties of a Nash equilibrium, with the addition that each player can guarantee a lower bound on his payoff without knowing the actions of other players. In other words, it replicates the properties of a Nash equilibrium with maximin strategies, as discussed in Pruzhansky \cite{pruzhansky2011some}. This is crucial in the context of a game option as it allows for super-replication. Furthermore, as shown in Proposition \ref{propoptivalue}, all optimal equilibria achieve the same value. 

\begin{definition}[Minimax, Maximin, Value] \label{defminimax}\ 
\begin{itemize}
	\item The \emph{maximin value} of player $k$ is the maximum payoff he can guarantee.
	\[
	\underline V_k=\max_{s_k\in\Strat_k}\min_{s_{-k}\in\Strat_{-k}} V_k\left(\big[s_k, s_{-k}\big]\right)
	\]
	\item The \emph{minimax value} of player $k$ is the lowest payoff that the other players can force upon him.
	\[
	\overline V_k=\min_{s_{-k}\in\Strat_{-k}}\max_{s_k\in\Strat_k} V_k\left(\big[s_k, s_{-k}\big]\right)
	\]
	\item In general,	$\overline V_k \geq \underline V_k$. If equality is achieved, then $V_k^*=\overline V_k = \underline V_k$ is the \emph{value} of the game for player $k$.
\end{itemize}
\end{definition}

In general, a game may not have a value. But the existence of an optimal equilibrium guarantees one.

\begin{proposition}\label{propoptivalue}
Let $s^*$ be any optimal equilibrium. Since 
\begin{gather*}
V_k\left(\big[s_k^*, s_{-k}^*\big]\right) = \max_{s_k\in\Strat_k} V_k\left(\big[s_k, s_{-k}^*\big]\right) \geq \overline V_k \\
\geq \underline V_k\geq \min_{s_{-k}\in\Strat_{-k}} V_k\left(\big[s_k^*, s_{-k}\big]\right)= V_k\left(\big[s_k^*, s_{-k}^*\big]\right),
\end{gather*}
all expressions are equal. Hence every optimal equilibrium attains the value of the game for all players
\[
\bm{V}(s^*)=\bm{V}^*=\big[V_1^*,\ldots,V_m^*\big].
\]
\end{proposition}

It is easily shown that any pure strategy Nash equilibrium is still a Nash equilibrium in the mixed strategy game. The same argument can be readily applied to show that any pure strategy optimal equilibrium is also an optimal equilibrium if mixed strategies are allowed. Hence the value of a pure strategy game is also the value of its mixed strategy extension.

In two person zero-sum games, optimal equilibria are equivalent to Nash equilibria. In fact Proposition \ref{propoptivalue} simply reduces to John von Neumann's minimax theorem \cite{v1928theorie}. The theorem states that in any two person, zero-sum game with finite strategies, there exists a unique payoff where mixed strategy Nash equilibria are achieved. This unique payoff is also the value of the game.

In multi-player zero-sum games, the payoff of any particular player is not sufficient to determine the other individual payoffs. Nash equilibria are not necessarily optimal equilibria, and they may not achieve the same payoff. However, a result similar to the minimax theorem exists if the zero-sum condition is strengthened to \emph{weakly unilaterally competitive}, as described in Kats and Thisse \cite{kats1992unilaterally} and De Wolf \cite{de1999optimal}.

\begin{definition}[Weakly Unilaterally Competitive]\label{defwuc}
A game is said to be \emph{weakly unilaterally competitive} (or WUC) if for any $k, l \in\Mset$:
\begin{align*}
V_k(\big[s_k,s_{-k}\big]) > V_k(\big[s_k',s_{-k}\big]) &\Longrightarrow V_l(\big[s_k,s_{-k}\big]) \leq V_l(\big[s_k',s_{-k}\big]),\\
V_k(\big[s_k,s_{-k}\big]) = V_k(\big[s_k',s_{-k}\big]) &\Longrightarrow V_l(\big[s_k,s_{-k}\big]) = V_l(\big[s_k',s_{-k}\big])
\end{align*}
for all $s_k,s_k'\in\Strat_k$ and $s_{-k}\in\Strat_{-k}$. 
\end{definition}

WUC explicitly quantifies the concept of competitiveness. If a player deviates from a strategy profile, any changes to his payoff is opposite in sign to the changes of other payoffs. Both Kats and Thisse \cite{kats1992unilaterally} and De Wolf \cite{de1999optimal} proved the following result, an analogue of the minimax theorem in multi-player, WUC settings.
\begin{proposition}\label{thmwucequi}
In a WUC game, any Nash equilibrium is also an optimal equilibrium.
\end{proposition}
WUC is not a necessary condition for the existence of optimal equilibria, but it eliminates the possibilities of Nash equilibria achieving multiple values. It is a desirable condition for the construction of our game.

Finally, this paper will only focus on pure strategy games. As shown in Ferenstein \cite{ferenstein2007randomized}, a stopping game with mixed or randomised strategies (in this case, randomised stopping times) can be reformulated as a stopping game with pure strategies with an appropriate filtration enlargement. Furthermore, in the context of evaluating game options, it is not practical to implement mixed or randomised strategies during payoff replication.

\section{Single Period Deterministic Games} \label{secdeterm}

Throughout this paper, the game option terminology of ``exercise'' will be utilised when referring to the stopping or quitting of the game by the players.
The corresponding payoff from doing so will be called ``exercise payoffs''.

Before proposing a multi-player variant, it is useful to recall the mechanism of a two person game option, as defined in Kifer \cite{kifer2000game}.
Essentially, the game option is a contract where the buyer can exercise the option at any time $t$ for a payoff $X_t$, while the seller
can cancel (or also ``exercise'') the option at any time $t$ for a cancellation fee of $Y_t$. If no one does anything, the contract
will expire at time $T$ with the buyer receiving $X_T$ from the seller.

\begin{remark}\label{remtwoplayer}
It is common to postulate that the inequality
$X_t\leq Y_t$ holds for every $t$. In other words, the cancellation fee should always at least as great as the exercise payoff. This circumvents the need to deal with simultaneous exercise for the purpose of valuation.  When the buyer exercises, it will cost the seller at least as much if he also cancels.
Similarly when the seller cancels, the buyer can only lose by exercising. If the players are exercising optimally, simultaneous
exercise only occurs when the equality $X_t=Y_t$ is true, in which case the payoff is still well defined.

The assumption of $X_t\leq Y_t$ can however be removed if we simply add another rule. If exercise and cancellation occur simultaneously, then the buyer and the seller simply receive $X_t$ and $Y_t$ respectively. This offers a cleaner mechanism which can be used in non zero-sum games, yet still produces the same value if $X_t\leq Y_t$ holds.
\end{remark}

When there are more than two players, symmetry needs to be introduced between the buyer and the seller. The
cancellation fee process can be interpreted as a negative exercise payoff process for the seller. So each player
 $k$ has his own exercise payoff amount $X_k$. And in the event that no player exercises, each player should receive a terminal payoff $P_k$. In the two player version, the terminal payoffs $P_1$ and $P_2$ correspond to $X_T$ and $-X_T$.

\begin{remark}\label{remcostexercise}
There are various ways to generalise the exercise mechanism. In the two player version, when one person exercises, the entire effect (or ``cost'') of that action, whether positive of negative, is paid by the other, non-exercising player. This paper will focus on a natural extension of that for multiple players, in which the effect of exercise is reflected in the payoffs of the non-exercising players, according to some weight function.
\end{remark}

We begin by setting up a single period deterministic game with $m$ players, where exercising is only allowed at one predetermined time.

\begin{definition}\label{defgameoption}
A \emph{single period deterministic multi-player game} $\Game$, with players indexed by $\Mset=\{1,2,\ldots,m\}$, is specified by the following:
\begin{itemize}
	\item The vector $\bm{X}=[X_1,\ldots,X_m]$, where $X_{k}$ is the amount received by player $k$ if he \emph{exercises} at time $0$.
	\item The vector $\bm{P}=[P_1,\ldots,P_m]$, where $P_k$ is the amount received by player $k$ if \emph{no player exercises} at time 0;
\end{itemize}
The rules of the game are:
\begin{enumerate}
	\item The strategy $s_k\in\Strat_k$ of player $k$ specifies whether player $k$ exercises, where $\Strat_k = \left\{0, 1\right\}$ is the space of strategies. In particular, $s_k=0$ means that player $k$ exercises at time 0, whereas $s_k=1$ means that player $k$ does not exercise.
	\item Given a strategy profile $s\in\Strat=\prod_{i\in\Mset} \Strat_i$, the \emph{exercise set}, denoted by $\Eset(s)$, is the set of exercising players.
	\item For each strategy profile $s$, the outcomes of the game $\Game$ are represented by the \emph{payoff} vector $\bm{V}(s)=[V_1 (s),\ldots,V_m (s)]$, where $V_k(s)$ is the payoff received by player $k$ if a strategy profile $s$ is carried out. It equals
\begin{align*}
V_k(s)=
\begin{cases}
 X_{k}, & k\in\Eset(s),\\
 P_k - w_k(\Eset(s)) D(s), & k\in \Mset\setminus\Eset(s),
\end{cases}
\end{align*}
where 
\begin{align*}
D(s) = \sum_{i\in \Eset(s)} (X_{i} - P_i)
\end{align*}
is the \emph{difference due to exercise} and $w_k(\Eset(s))$ is a \emph{weight} function which will be chosen in subsection \ref{secweight} (cf. Definition \ref{assumpw3}).
\end{enumerate}
\end{definition}

The payoffs $V_k(s)$ are linear functions of $X_i$ and $P_i$. Most results of the deterministic case can hence be easily extended to stochastic settings, which will be covered in subsection \ref{secstoch}. The difference due to exercise $D(s)$ represents the effect of exercise described in Remark \ref{remcostexercise}, and they negatively affect the payoff of the non-exercising players if the weights are positive. As seen later in Theorem \ref{thmweightswuc}, the positivity of weights will be an important condition for a WUC game.

If $\sum_{i\notin\Eset} w_i(\Eset)=1$ is satisfied for every $\Eset\subset\Mset$, then the sum of all payoffs is either $\sum_{i\in\Mset} X_i$ (everyone exercises) or $\sum_{i\in\Mset} P_i$ (at least one player doesn't exercise). This correspond to the property of the two player zero-sum game described in Remark \ref{remtwoplayer}. In fact, the two player game can be recovered by setting $w_1(\{2\})=w_2(\{1\})=1$.

\subsection{Choice of Weights}\label{secweight}

In order to proceed further, we need to be more explicit about the way
in which the weights are specified. In general $w_k(\Eset)$ is defined for non-empty subsets $\Eset\subset\Mset$, $\Eset\neq\Mset$ and $k\notin\Eset$. Consider the class of games with fixed weights $w_k(\Eset)$ but with all possible choices of $\bm{X}$ and $\bm{P}$. The goal of this section is to find weights achieving two \emph{conditions}:
\begin{enumerate}
	\item Every game in the class is WUC.
	\item Every game in the class has at least one equilibrium in pure strategies.
\end{enumerate}
The motivation is to separate the exercise mechanism, driven by $w_k(\Eset)$, from the payoff parameters $\bm{X}$ and $\bm{P}$. Thus the game retains nice properties irrespective of the choice of $\bm{X}$ and $\bm{P}$. This is especially useful in stochastic cases and game options.

Furthermore, we will assume that $w_k(\Eset)\neq 0$ for all non-empty subset $\Eset\subset\Mset$, $\Eset\neq\Mset$ and $k\notin\Eset$.
This assumption ensures that the decision of any exercising player will always affect the payoff of a non-exercising player. It also eliminates various degenerate cases. The WUC condition further refines the restrictions on the weights, as shown below.

\begin{proposition}\label{thmweightswuc}
Fixing the weights $w_k(\Eset)\neq 0$, the game $\Game$ is WUC for all choices $\bm{X}$ and $\bm{P}$ if and only if the weights can be written in the following form:
\begin{align}\label{eq1}
w_k(\Eset) = \frac{a_k}{1-\sum_{i\in\Eset}a_i},\quad\text{where } a_k > 0 \text{ and } \sum_{i\neq k} a_i < 1 \text{ for all }k.
\end{align}
\end{proposition}
\begin{proof}See Appendix \ref{secappendix1}.
\end{proof}
In \eqref{eq1}, the term $a_k$ indicates how much player $k$ is affected by the exercise decisions of others. The relative size of $a_k$ determines the relative size of the weights $w_k(\Eset)$, used to redistributing $D(s)$. By Proposition \ref{thmweightswuc}, the game with weights defined by \eqref{eq1} is always WUC, fulfilling the first condition. For the second condition of always having a pure strategy Nash equilibrium, further restrictions are needed.

\begin{proposition}\label{propalwaysnash}
Using weights defined by \eqref{eq1}, if $\sum_{i\in\Mset} a_i > 1$, then there exist $\bm{X}$ and $\bm{P}$ such that no equilibrium exists in pure strategies. In particular, $\bm{X}=[0,0,\ldots,0]$ and $\bm{P}=[1,-1,0,\ldots,0]$.
\end{proposition}
\begin{proof}See Appendix \ref{secappendix1}.
\end{proof}
Combining Propositions \ref{thmweightswuc} and \ref{propalwaysnash}, we arrive at the following definition.
\begin{definition}[Weights]\label{assumpw3}
For each non-empty subset $\Eset\subset\Mset$, $\Eset\neq\Mset$ and $k\notin\Eset$, define the \emph{weight} $w_k(\Eset)$ by
\[
w_k(\Eset)=\frac{a_k}{1-\sum_{i\in\Eset}a_i}
\]
where $a_1,\ldots,a_m$ are real numbers satisfying $a_k>0$ and $\sum_{k\in\Mset} a_k\leq 1$. In particular, $0<w_k(\Eset)\leq 1$.
\end{definition}
Definition \ref{assumpw3} will be used as the definition of weights for the remainder of the paper. As shown later in Theorem \ref{thmsoln}, the game with these weights always has a value, achieved by all Nash and optimal equilibria.

\subsection{Existence of Value}

With the weights determined by Definition \ref{assumpw3}, the game in Definition \ref{defgameoption} is now fully specified. The main result of this subsection, as well as this paper, is to show that the game always has a value. Recall that the uniqueness of the value was already established in Definition \ref{defminimax}.  

\begin{theorem}\label{thmsoln}
The game $\Game$ from Definition \ref{defgameoption} has at least one optimal equilibrium. Hence it has a value.
\end{theorem}
Before proving Theorem \ref{thmsoln}, a couple of preliminary results are required. Firstly, Lemma \ref{lemsubgame} shows that the subgames of $\Game$ retain the same structure as $\Game$. This is a consequence of Definition \ref{assumpw3}.

In the game $\Game$, consider a subset $\Eset \subset \Mset$ and assume that every player in $\Eset$ exercises at time 0, while the players from $\Mset'=\Mset\setminus\Eset$ are still free to make choices. The possible outcomes of the game for the players from $\Mset'$ define a subgame of $\Game$, which we denote by $\Game'$. It is possible to show that the $\Game'$ behaves similar to $\Game$, but with modified parameters.
\begin{lemma}[Subgame]
\label{lemsubgame}
The subgame $\Game'$ is equivalent to a game $\Game_{\Mset'}$ specified by the following:
\begin{itemize}
\item Set of players $\Mset'=\Mset\setminus\Eset$;
\item $X_k'=X_k$, for $k\in\Mset'$;
\item $P_k'=P_k-w_k(\Eset)\sum_{i\in\Eset}(X_i-P_i)$, for $k\in\Mset'$;
\item Weights defined by $a_k'=w_k(\Eset)=\frac{a_k}{1-\sum_{i\in\Eset}a_i}$, for $k\in\Mset'$; and
\item Strategy profile $s'=[s_k',\ k\in\Mset']\in \Strat_{\Mset'}$.
\end{itemize}
\end{lemma}

\begin{proof}See Appendix \ref{secappendix3}.
\end{proof}
The property of the subgames proven in Lemma \ref{lemsubgame} is useful because it allows for the use of induction on the number of players. Furthermore it allows us to extend any equilibrium results to subgame perfect equilibrium results (for example, in section \ref{secquitgames}). 

We also require Lemma \ref{lemposdiff}, which asserts that the difference due to exercise has to be positive under any Nash equilibrium.

\begin{lemma} \label{lemposdiff}
If $s^*$ is an equilibrium then $D(s^*) =  \sum_{i\in \Eset(s^*)} (X_{i} - P_i)  \geq 0$.
\end{lemma}
\begin{proof}Intuitively, if $\sum_{i\in \Eset(s^*)} (X_{i} - P_i) < 0$, then there exists a player $k\in\Eset{s^*}$ who can do better by not exercising. For details see Appendix \ref{secappendix3}.
\end{proof}

Now we are ready for the proof of Theorem \ref{thmsoln}.
\begin{proof}[Theorem \ref{thmsoln}]
By Proposition \ref{thmwucequi}, in a WUC game, all Nash equilibria are optimal equilibria and attain the same value. So it is sufficient to construct a Nash equilibrium in pure strategies.

We proceed by induction on the number of players. Note that the game is still well-defined as a single person game when $m=1$, and all single person games have at least one equilibrium. In particular, $s^*=[1]$ if $P_1>X_1$ or $s^*=[0]$ if $P_1\leq X_1$.

Consider $m\geq 2$. If $P_i > X_i$ for all $i$, then $s^*=\bm{1}$ is an equilibrium. If $P_k \leq X_k$ for some $k$, consider the $m-1$ player subgame $\Game_{\{-k\}}$. Let $s'$ be an equilibrium of $\Game_{\{-k\}}$, which exists by the induction hypothesis. Consider the strategy profile $s^*=[1, s_{-k}=s']$, we will show it is an equilibrium of $\Game$.

By construction, $s'$ is an equilibrium of $\Game_{\{-k\}}$, so any player $i\neq k$ cannot improve his payoff by changing strategy. Hence it is sufficient to check that player $k$ cannot improve by not exercising, or $V_k(s^*)=X_k\geq V_k(s)=P_k-w_k(\Eset)D(s)$ where $s=[0, s_{-k}=s']$ and $\Eset(s')=\Eset(s)=\Eset$. Now write $D(s)$ in terms of the subgame $\Game_{\{-k\}}$ variables $P_i'$ and $D'(s')$ (as defined in Lemma \ref{lemsubgame})
\begin{align}\label{eq10}
D(s)=\sum_{i\in\Eset}(X_i-P_i')+(P_i'-P_i)=D'(s')-\sum_{i\in\Eset}w_i(\{k\})(X_k-P_k).
\end{align}
Substituting \eqref{eq10} back, we want the following expression to be non-negative,
\begin{align}\label{eq11}
&X_k-(P_k-w_k(\Eset)D(s))\nonumber\\
={}&X_k-P_k+w_k(\Eset)D'(s')-w_k(\Eset)(X_k-P_k)\sum_{i\in\Eset}w_i(\{k\})\nonumber\\
={}&w_k(\Eset)D'(s')+(X_k-P_k)\bigg(1-w_k(\Eset)\sum_{i\in\Eset}w_i(\{k\})\bigg).
\end{align}
Applying Lemma \ref{lemposdiff} to the subgame $\Game_{\{k\}}$, $D'(s')\geq 0$. Also recall $w_k(\Eset)\geq 0, (X_k-P_k) \geq 0$ by assumptions. It remains to check the last term of \eqref{eq11}, which can be written as
\begin{align}\label{eq12}
1-w_k(\Eset)\sum_{i\in\Eset}w_i(\{k\})=1-\frac{a_k}{1-\sum_{i\in\Eset}a_i}\frac{\sum_{i\in\Eset}a_i}{1-a_k}=\frac{1-a_k-\sum_{i\in\Eset}a_i}{\left(1-\sum_{i\in\Eset}a_i\right)(1-a_k)}.
\end{align}
Finally, \eqref{eq12} is non-negative because $a_k+\sum_{i\in\Eset}a_i\leq \sum_{i\in\Mset}a_i \leq 1$. So we indeed have $X_k\geq V_k(s)$. Therefore $s^*$ is an equilibrium and the induction is complete.
\end{proof}

An algorithm of finding an equilibrium $s^*$ follows from the proof of Theorem \ref{thmsoln}, by repeatedly identifying players with $P_i\leq X_i$ as exercising players, and reducing the problem to a smaller case. The equilibrium and its value are found within $m-1$ iterations.

\section{Construction of Value} \label{secprojection}

In section \ref{secdeterm}, a multi-player game $\Game$ is introduced to model a multilateral contract, where the payoffs have an implicit competitive structure. Theorem \ref{thmsoln} established that $\Game$ has a value, using an inductive argument. In this section, Theorem \ref{thmsoln3} will give an explicit construction of the value. The construction is motivated by Guo and Rutkowski \cite{guo2012zero}, which discussed a special case of $\Game$ which is zero-sum, and expressed the value in terms of a projection. Subsection \ref{secprojconstsum} will briefly revisit these results, then subsection \ref{secprojgen} will extend them to the general case.

We begin by identifying the payoffs as as vectors in $\R^m$ and introducing the following notations.
\begin{definition}[Hyperplane and Orthant]\label{defhypeandorthant}
For any subset $\Eset\subseteq\Mset$, define the \emph{hyperplane} $\H_\Eset$ as
\[
\H_\Eset = \bigg\{ \bm{x} \in \R^m :\,  x_i = X_{i},\ \forall\, i\in \Eset\bigg\}.
\]
In particular, $\H_\emptyset=\R^m$ and $\H_\Mset = \bm{X}$. The hyperplane $\H_\Eset$ contains all the possible payoffs if all players in $\Eset$ exercise. Also define the \emph{orthant} $\O$ as
\[
\O = \bigg\{ \bm{x} \in \R^m :\,  x_i\geq X_{i}, \, 1\leq i \leq m \bigg\}.
\]
The boundary of $\O$ are the hyperplanes $\H_{\{i\}}$.
\end{definition}

\begin{proposition}\label{proporthant}
The value $\bm{V}^*$ of the game $\Game$ must lie inside the orthant $\O$.
\end{proposition}
\begin{proof}
Given any equilibrium $s^*$, the payoff of each player should be at least as great as his exercise payoff, or $V_k(s^*)\geq X_k$ for all $k\in\Mset$. Hence $\bm{V}^*\in\O$.
\end{proof}

\begin{remark}\label{defprojection}
Given a normed vector space $\R^m$, for any vector $\bm{P}$ and any closed convex set $\K$, denote by $\proj{\bm{P}}{\K}$ the projection of $\bm{P}$ onto $\K$. In other words, $\proj{\bm{P}}{\K}$ is a vector satisfying $ \proj{\bm{P}}{\K}\in\K $ and
\[
\norm{\proj{\bm{P}}{\K}-\bm{P}}\leq \norm{\bm{Q}-\bm{P} },\quad \forall\, \bm{Q}\in\K.
\]
For the normed spaces discussed in the following subsections, the existence and uniqueness of the projection are well-known.
\end{remark}

\subsection{Zero and Constant Sum Cases}\label{secprojconstsum}

This subsection will only present the main definitions and results without proof. A detailed treatment can be found in Guo and Rutkowski \cite{guo2012zero}. Note that analogous results for the non zero-sum cases will be presented in subsection \ref{secprojgen} (cf. Proposition \ref{thmexproj1}, Theorem \ref{thmsoln3} and Remark \ref{remgamelimit}).

Only in this subsection, add the following constraints to Definition \ref{assumpw3}, the definition of weights:
\begin{align}\label{eq13}
\sum_{i\in\Mset} a_i=1\quad \Longrightarrow\quad \sum_{i\notin\Eset} w_i(\Eset)=\frac{\sum_{i\notin\Eset}a_i}{1-\sum_{j\in\Eset}a_j}=1\quad \forall\,\Eset\subset\Mset, \Eset\neq\emptyset.
\end{align}
Denote the resulting game by $\Game^0$. If $\Eset(s)=\Mset$, then $\sum_{i\in\Mset} V_i(s)=\sum_{i\in\Mset} X_i$. Otherwise,
\begin{align*}
\sum_{i\in\Mset} V_i(s)&= \sum_{i\in\Eset} X_i + \sum_{i\notin\Eset}\bigg(P_i-w_i(\Eset)\sum_{j\in\Eset}(X_j-P_j)\bigg)\\
 &= \sum_{i\in\Eset} X_i + \sum_{i\notin\Eset}P_i-\sum_{j\in\Eset}(X_j-P_j)=\sum_{i\in\Mset} P_i.
\end{align*}
So the payoffs of $\Game^0$ have a constant sum, except when everyone exercises. All possible payoffs except one lie on the hyperplane
\[
\H^0 = \bigg\{  \bm{x} \in \R^m :\, \sum_{i=1}^m x_i = \sum_{i=1}^m P_i \bigg\}.
\]
In particular, $\Game^0$ is almost zero sum if $\sum_{i\in\Mset} P_i=0$. Next, for any proper subset $\Eset\subset\Mset$, define the hyperplane
\[
\H_\Eset^0 = \H_\Eset\cap\H^0 = \bigg\{ \bm{x} \in \R^m :\,  x_i = X_{i}\ \forall\,i\in \Eset \ \mbox{and}\ \sum_{i=1}^m  x_i = \sum_{i=1}^m P_i \bigg\}.
\]
In particular $\H_\emptyset^0=\H^0$. This definition is not adequate when $\Eset=\Mset$, because $\H_\Mset\cap\H^0=\bm{X}\cap\H^0$ is empty unless $\sum_{i\in\Mset} X_{i} = \sum_{i\in\Mset} P_i$. But for completeness we further define $\H_\Mset^0=\H_\Mset=\bm{X}$.

\begin{definition}[Inner Product and Norm]\label{definner1}
Endow $\R^{m}$ with the following \emph{inner product} and \emph{norm}
\[
\langle\bm{x},\bm{y}\rangle^0 = \sum_{i=1}^{m} \left( \frac{x_iy_i}{a_i}\right),\qquad\norm{\bm{x}}^0 = \left(\sum_{i=1}^m\frac{x_i^2}{a_i}\right)^{\frac{1}{2}}.
\]
\end{definition}
The choice of norm in Definition \ref{definner1} will be justified by Proposition \ref{thmexproj0}, which gives an alternative yet elegant way of representing and computing the payoff vector, provided that the exercise set is known.

\begin{proposition}[Payoff as Projection]
\label{thmexproj0}
For any strategy profile $s \in \Strat$, the payoff vector $\bm{V}(s)$ equals
\[
\bm{V}(s) = \proj{\bm{P}}{\H_{\Eset(s)}^0}
\]
where the projection $\pi$ is taken under the norm $\norm{\cdot}^0$.
\end{proposition}

Consider the simplex $\S$ given by the formula
\[
\S = \O\cap\H^0  =\bigg\{ \bm{x} \in \R^m :\,  x_i\geq X_{i}, \, 1\leq i \leq m \ \mbox{and}\ \sum\limits_{i=1}^m x_i = \sum_{i=1}^m P_i \bigg\}.
\]
There are a few possibilities:
\begin{itemize}
\item If $\sum_{i\in\Mset}X_i > \sum_{i\in\Mset}P_i $, $\S$ is empty.;
\item If $\sum_{i\in\Mset}X_i = \sum_{i\in\Mset}P_i $, $\S=\H_\Mset^0=\bm{X}$ is a single point; 
\item If $\sum_{i\in\Mset}X_i < \sum_{i\in\Mset}P_i $, $\S$ is an $m-1$ dimensional simplex, whose faces are $\S\cap\H_\Aset^0$ as $\Aset$ varies over the subsets of $\Mset$.
\end{itemize}
As mentioned in Proposition \ref{proporthant}, the value $\bm{V}^*$ must lie inside $\O$. Since the all payoffs of $\Game^0$ but one lie on $\H^0$, we expect $\bm{V}^*$ to lie in $\S$ (if it is non-empty). The following theorem shows that $\bm{V}^*$ is in fact the projection of $\bm{P}$ onto $\S$.

\begin{theorem}[Value by Projection]
\label{thmsoln2}
In game $\Game^0$, if $\sum_{i\in\Mset}X_i \leq  \sum_{i\in\Mset}P_i$, the value $\bm{V}^*$ is
\begin{align*}
\bm{V}^*= \bm{V} (s^*) =  \proj{\bm{P}}{\S}
\end{align*}
where the projection $\pi$ is taken under the norm $\norm{\cdot}^0$ from Definition \ref{definner1}. An optimal equilibrium $s^*=[s_1^*,\ldots,s_m^*]$ is given by
\[
s_i^*=0\quad\Longleftrightarrow\quad\big[\proj{\bm{P}}{\S}\big]_i=X_i.
\]
If $\sum_{i\in\Mset}X_i >  \sum_{i\in\Mset}P_i$, the value is $\bm{V}^*= \bm{V} (s^*)= \bm{X}$ 
with an optimal equilibrium given by $s^*=\bm{0}$.
\end{theorem}
The two cases in Theorem \ref{thmsoln2} resemble the phenomena described in Remark \ref{remtwoplayer} for two player game options. In the first case, $\sum_{i\in\Mset}X_i \leq  \sum_{i\in\Mset}P_i$ is analogous to the condition $X_t\leq Y_t$. On the other hand, if $\sum_{i\in\Mset}X_i >  \sum_{i\in\Mset}P_i$, the game becomes degenerate and every player exercises. The effect is similar to the case of $X_t > Y_t$ in two player game options.

\subsection{Non Zero-Sum Cases}\label{secprojgen}

This subsection extends the projection representation to cases where $\Game$ is not constant (or zero) sum, or $\sum_{i\in\Mset}a_i<1$. The key idea is to construct an equivalent constant (or zero) sum game by introducing a {dummy} player.

Given an $m$ player game $\Game$ with $\sum_{i\in\Mset}a_i<1$, construct an $m+1$ player game $\Game^0$ by adding a \emph{dummy player} with the following attributes:
\begin{itemize}
	\item No exercising allowed;
	\item $P_{m+1}$ is arbitrary, but for the sake of simplicity set $P_{m+1}=-\sum_{i\in\Mset} P_i$, so $\Game^0$ is zero-sum;
	\item Weights $w_{m+1}(\Eset)=\frac{a_{m+1}}{1-\sum_{i\in\Eset} a_i}$ where $a_{m+1} = 1-\sum_{i\in\Mset}a_i>0$;
	\item Payoff function $V_{m+1}(s)=P_{m+1}-w_{m+1}(\Eset(s))\sum_{i\in\Eset(s)}(X_i-P_i)$.
\end{itemize}
The space of strategy profiles in $\Game^0$ is the same as $\Strat$ from $\Game$, because player $m+1$ cannot make any choices. The payoffs of players in $\Mset$ do not change in $\Game^0$. Any Nash/optimal equilibrium for $\Game^0$ is also a Nash/optimal equilibrium of $\Game$ and vice versa. Finally the value of $\Game$ equals the value of $\Game^0$, after restricting to the first $m$ coordinates.

Since $\Game^0$ is zero-sum, the notations from subsection \ref{secprojconstsum} can also be used here. Furthermore $\Game^0$ has the advantage of the dummy player not being able to exercise, so the game is always zero-sum.
In other words, $V_{m+1}(s)+\sum_{i\in\Mset}V_i(s)=0$ for all $s\in\Strat$. Hence the space of payoff vectors for $\Game^0$ is the hyperplane $\H^0=\{\bm{x}: \sum_{i=1}^{m+1}x_i=0\}\subset\R^{m+1}$, endowed with the inner product $\langle\cdot,\cdot\rangle^0$ from Definition \ref{definner1},
\begin{align}\label{eq15}
\langle\bm{x},\bm{y}\rangle^0 = \sum_{i=1}^m \left( \frac{x_iy_i}{a_i}\right) + \frac{\left(-\sum_{i=1}^m x_i \right)\left(-\sum_{i=1}^m y_i\right)}{a_{m+1}}.
\end{align}
Rewriting $a_{m+1}$ in terms of $a_1,\ldots,a_m$, \eqref{eq15} motivates the following inner product and norm for $\R^m$, the space of payoffs for $\Game$.

\begin{definition}[Inner Product and Norm]\label{definner2}
Endow $\R^{m}$ with the following \emph{inner product} and \emph{norm}
\begin{align*}
\langle\bm{x},\bm{y}\rangle &= \sum_{i=1}^m \left( \frac{x_iy_i}{a_i}\right) + \frac{\left(\sum_{i=1}^m x_i \right)\left(\sum_{i=1}^m y_i\right)}{1-\sum_{i=1}^m a_i},\\
\norm{\bm{x}} &= \left(\sum_{i=1}^m \left( \frac{x_i^2}{a_i}\right)+ \frac{\left(\sum_{i=1}^m x_i \right)^2}{1-\sum_{i=1}^m a_i}\right)^{\frac{1}{2}}.
\end{align*}
\end{definition}

Consider the isometry $\phi: \H^0\rightarrow\R^m$ which simply discards the $(m+1)$-th coordinate. It maps  $\H_\Aset^0=\H_\Aset\cap\H^0$ and $\S=\O\cap\H^0$ to $\H_\Aset$ and $\O$ respectively. Equipped with the new inner product $\langle\cdot,\cdot\rangle$, we can discard the extra player, and return to $\R^m$ and $\Game$. Proposition \ref{thmexproj0} and Theorem \ref{thmsoln2} can now be analogously formulated for the general case, as Proposition \ref{thmexproj1} and Theorem \ref{thmsoln3}. The proofs are in Appendix \ref{secappendix2}.

\begin{proposition}[Payoff as Projection]
\label{thmexproj1}
For any strategy profile $s \in \Strat$, the payoff vector $\bm{V}(s)$ equals
\[
\bm{V}(s) = \proj{\bm{P}}{\H_{\Eset(s)}}
\]
where the projection $\pi$ is taken under the norm $\norm{\cdot}$.
\end{proposition}

\begin{theorem}[Unique Value by Projection]
\label{thmsoln3}
In the game $\Game$, the value is given by
\begin{align*}
\bm{V}^*= \bm{V} (s^*) = [V_1(s^*),\ldots,V_m(s^*)] =  \proj{\bm{P}}{\O}
\end{align*}
where the projection $\pi$ is taken under the norm $\norm{\cdot}$. An optimal equilibrium $s^*=[s_1^*,\ldots,s_m^*]$ is given by
\[
s_i^*=0\quad\Longleftrightarrow\quad\big[\proj{\bm{P}}{\O}\big]_i=X_i.
\]
\end{theorem}

The existence of the value was first established in Theorem \ref{thmsoln}. Theorem \ref{thmsoln3} reaffirms that result by providing an explicit representation using a concise notation. The chosen optimal equilibrium $s^*$ is determined by identifying the exercise set $\Eset(s^*)$ from the hyperplanes used in projection $\proj{\bm{P}}{\O}=\proj{\bm{P}}{\H_{\Eset(s^*)}}$. 

\begin{remark}\label{remgamelimit}
If the original game $\Game$ is already constant (or zero) sum, then the introduction of player $m+1$ is problematic. Division by zero occurs because $a_{m+1}=1-\sum_{i\in\Mset} a_i=0$, so $\Game^0, \langle\cdot\rangle^0$ and $\norm{\cdot}^0 $ from subsection \ref{secprojconstsum} are no longer well defined. However, Proposition \ref{thmexproj1} and Theorem \ref{thmsoln3} can still be applied to the constant (or zero) sum case if the following conventions are adopted.

Consider the $m+1$ player game $\Game^\epsilon$, where $a_1,\ldots,a_m$ and $a_{m+1}$ are replaced by
\[
a_i^\epsilon=a_i-\frac{\epsilon}{m}, i\in\Mset;\quad a_{m+1}^\epsilon=1-\sum_{i\in\Mset} a_i^\epsilon=\epsilon.
\]
Take $\Game^0$ to be the limit of $\Game^\epsilon$ as $\epsilon\rightarrow 0$. Since for any fixed $k\in\Mset$ and any fixed $s\in\Strat$, the $\Game^\epsilon$ payoff $V^\epsilon_k(s)$ is continuous (linear, in fact) in $\epsilon$, the limit $V^0_k(s)$ is indeed the desired $\Game^0$ payoff. Furthermore, 
$\underline V_k^\epsilon=\max_{s_k}\min_{s_{-k}} V_k^\epsilon\left(\big[s_k, s_{-k}\big]\right)$ and $\overline V_k^\epsilon=\min_{s_{-k}}\max_{s_k} V_k^\epsilon\left(\big[s_k, s_{-k}\big]\right)$ are also continuous in $\epsilon$, so the value of $\Game^\epsilon$ converges to the value of $\Game^0$ as well.

Once again, the payoffs and value of the $m$ player game $\Game$ are obtained from $\Game^0$ after discarding the $(m+1)$-th coordinate. Although $\norm{\cdot}$ does not exist, Proposition \ref{thmexproj1} and Theorem \ref{thmsoln3} can still be recovered by redefining the projection $\pi$. We formally adopt the following convention for the remainder of the paper.

\begin{definition}[Projection]\label{defprojconvention}
When $\sum_{i\in\Mset}a_i<1$, then define $\pi$ as the projection (see Remark \ref{defprojection}) under $\norm{\cdot}$ from Definition \ref{definner2}. In the case of $\sum_{i\in\Mset}a_i=1$, define $\pi$ for $\H_\Eset, \Eset\subseteq\Mset$ and $\O$ as follows,
\[
\proj{\bm{P}}{\H_\Eset}=\lim_{\epsilon\rightarrow 0} \pi_{\H_\Eset}^\epsilon(\bm{P}),\quad \proj{\bm{P}}{\O}=\lim_{\epsilon\rightarrow 0} \pi_{\O}^\epsilon(\bm{P})
\]
where $\pi^\epsilon$ is the projection under the norm
\begin{align}\label{eq14}
\norm{\bm{x}}^\epsilon = \left(\sum_{i=1}^m \left( \frac{x_i^2}{a_i-\frac{\epsilon}{m}}\right)+ \frac{\left(\sum_{i=1}^m x_i \right)^2}{\epsilon}\right)^{\frac{1}{2}}.
\end{align}
\end{definition}

Intuitively the $\left(\sum_{i=1}^m x_i \right)^2/\epsilon$ term in \eqref{eq14} serves as a \emph{penalty} function, to keep the deviation of $\sum_{i=1}^m x_i$ to a minimum during projection. In subsection \ref{secprojconstsum}, the same effect is produced by restricting projections to the zero-sum hyperplane $\H^0$ (hence using $\H_\Aset^0$ and $\S$ instead of $\H_\Aset$ and $\O$). The penalty function allows for a cleaner representation consistent with the general case. The notation of $\proj{\bm{P}}{\O}$ in Theorem \ref{thmsoln3} also conveniently eliminates the need for cases in Theorem \ref{thmsoln2}.

\end{remark}

\section{Stochastic and Multiple Period Extensions} \label{secmultiperiod}

The goal of this section is to study various extensions of the single period deterministic game. We begin by presenting a straight forward single period stochastic version. For multiple periods, instead of only choosing between ``exercise'' and ``not exercise'', each player also have to decide ``when to exercise''. These games are rarely WUC (cf. Definition \ref{defwuc}), even if the single period building blocks are. Nevertheless, we attempt to identify generalisations where optimal equilibria and value still exist.

\subsection{Single Period Stochastic Games} \label{secstoch}

The stochastic game is very similar to the deterministic version from Definition \ref{defgameoption}, except the vectors $\bm{X}$ and $\bm{P}$ are not known at the time of exercise. All of the following definitions are taken under the probability space $(\Omega, \Filt, \P )$.

\begin{definition}[Stochastic Game]
\label{defstochgame}
A {\it single period stochastic multi-player game} $\Game$, with players indexed by $\Mset=\{1,2,\ldots,m\}$, is specified by the following:
\begin{itemize}
	\item The $\Filt$-measurable random vectors $\bm{P}=[P_1, \ldots, P_m]$, $\bm{X}=[X_1, \ldots, X_m]$;
	\item The weights $w_k(\Eset)=\frac{a_k}{1-\sum_{i\in\Eset}a_i}$ for $k\notin\Eset\subset\Mset$, where $a_i>0$ is deterministic and $\sum_{i\in\Mset} a_i \leq 1$.
\end{itemize}
The rules of the game are:
\begin{enumerate}
	\item The players are only allowed to exercise at time 0, when $\bm{P}$ and $\bm{X}$ may not be known exactly. The payoffs are distributed at time 1.
	\item The space of strategy profiles is $\Strat = \prod_{i\in\Mset} \Strat_i$, where $\Strat_k=\{0,1\}$ is the space of pure strategies for player $k$ with $s_k=0$ meaning player $k$ exercises at time 0. The exercise set $\Eset(s), s\in\Strat$ is the set of exercising players. 
	\item For each strategy profile $s$, the outcome of the game is the expected payoff vector $\bm{V}(s)=[V_1 (s),\ldots,V_m (s)]$, defined by
	\[
V_{k}(s)=\EP\big( X_{k} \I_{\{k\in\Eset(s)\}}+\widetilde X_{k}\I_{\{k\notin\Eset(s)\}} \big)
\]
where
\[
\widetilde X_{k}=P_k - w_k(\Eset(s))\sum_{k\in \Eset(s)}(X_{k} - P_k).
\]
\end{enumerate}
\end{definition}

By the linearity of the payoff function $V_k(s)$, the stochastic game is equivalent to a deterministic game $\Game'$ starting at time 0 with $\bm{P}'=\EP(\bm{P})$ and $\bm{X}'=\EP(\bm{X})$. Hence by Theorems \ref{thmsoln} and \ref{thmsoln3}, the value of $\Game$ is given by:

\begin{proposition}[Value by Projection]
\label{thmexpectedsoln}
The stochastic game $\Game$ has a unique value given by
\[
\bm{V}(s^*)=\proj{\EP \left(\bm{P}\right)}{\O }
\]
where $\O$ is the orthant defined by
\[
\O = \bigg\{\bm{x}\in\R^m:\, x_i\geq \EP\left(X_{i}\right), \ 1\leq i \leq m \bigg\}
\]
and the projection $\pi$ is given by Definition \ref{defprojconvention}. A possible optimal equilibrium $s^*=[s_1^*,\ldots,s_m^*]$ is given by
\[
s_i^*=0\quad\Longleftrightarrow\quad\big[\proj{\EP \left(\bm{P}\right)}{\O}\big]_i=X_i.
\]
\end{proposition}

\subsection{Multiple Period Stopping Games}

There are various ways of generalising two player game option to multiple players while keeping the underlying dynamics. This paper will consider a game similar in dynamic to the one studied in Solan and Vieille \cite{solan2001quitting}, which stops as soon as any subset of players chooses to stop. We will use a payoff function generalising the single period game from Definition \ref{defgameoption}, with the goal of constructing optimal equilibria in pure strategies. The game will be recursive, in the sense that each player chooses between an exercise payoff and the value of a shorter game from the next period onward. It is reminiscent of compound or nested financial options, and it reduces to the discrete time two player game contingent claim (cf. Kifer \cite{kifer2000game}) if $m=2$ and $a_1=a_2=1/2$. 

Using standard terminologies for stochastic processes, all of the following definitions are taken under the probability space $(\Omega, \Filt, \P )$ endowed with the filtration $\FF = \{\Filt_t: t=0,1,\ldots,T\}$, representing the possible exercise times.

\begin{definition}\label{defmultiperiodgame}
For $t=0,1,\ldots,T$, a \emph{multi-player stochastic stopping game} $\Game_t$, with players indexed by $\Mset=\{1,2,\ldots,m\}$, is defined on the time interval $[t,T]$, specified by the following inputs:
\begin{itemize}
	\item The $\FF$-adapted processes $\bm{X}_u=[X_{1,u}, \ldots, X_{m,u}]$, where $u=t,\ldots,T$;
	\item The weights $w_k(\Eset)=\frac{a_k}{1-\sum_{i\in\Eset}a_i}$ for $k\notin\Eset\subset\Mset$, where $a_i>0$ is deterministic and $\sum_{i\in\Mset} a_i \leq 1$.
\end{itemize}
The rules of the game are:
\begin{enumerate}
	\item Each player can exercise at any time in the interval $[t,T]$. The game stops as soon as anyone exercises. If no one exercises before time $T$, then everyone must exercise at time $T$.
	\item The strategy $s_{k,t}$ of player $k$ is a stopping time chosen from the space $\Strat_{k,t}$ of $\FF$ stopping times, in the interval $[t,T]$. The strategy profile $s_t=[s_{1,t},\ldots,s_{m,t}]\in\Strat_t$ is the $m$-tuple of stopping times. Denote $\widehat s_t = s_{1,t}\wedge \cdots \wedge s_{m,t}$ to be the minimal stopping time, also an $\FF$ stopping time. The exercise set $\Eset(s_t)=\{i\in\Mset: s_{i,t} = \widehat s_t\}$ is the random set of earliest stopping players. 
	\item For each strategy profile $s_t$, the outcome of the game is the expected payoff vector $\bm{V}_t(s_t)=[V_{1,t}(s_t),\ldots,V_{m,t}(s_t)]$, defined by
\begin{align}\label{eq20}
V_{k,t}(s_t)=\EP\big( X_{k,\widehat s_t} \I_{\{k\in\Eset(s_t)\}}+\widetilde X_{k,\widehat s_t}\I_{\{k\notin\Eset(s_t)\}} \,\big|\, \Filt_t \big)
\end{align}
where
\begin{align}\label{eq21}
\widetilde X_{k,\widehat s_t}=V_{k,\widehat s_t+1}^*  - w_k (\Eset(s_t))  \sum_{i \in \Eset(s_t)} \big( X_{i,\widehat s_t} - V_{i,\widehat s_t+1}^*\big),\quad \widehat s_t<T.
\end{align}
and $\bm{V}_{\widehat s_t+1}^*=\big[V_{1,\widehat s_t+1}^*,\ldots,V_{m,\widehat s_t+1}^*\big]$ is the value of the game $\Game_{\widehat s_t+1}$.

As the game is stopped at time $\widehat s_t$, the indicator functions in \eqref{eq20} separate the exercising players from the others. $X_{k,\widehat s_t}$ is the payoff for an exercising player while $\widetilde X_{k,\widehat s_t}$ is the payoff for a non-exercising player. The game $\Game_{\widehat s_t+1}$ can be considered as the continuation of the current game if it does not stop at time $\widehat s_t$. Note that in \eqref{eq21}, $\widetilde X_{k,\widehat s_t}$ is not defined for $\widehat s_t=T$, but this does not matter because if the game is only stopped at $T$, every player must exercise and receive $X_{k,T}$, not $\widetilde X_{k,T}$.
\end{enumerate}
\end{definition}
Definition \ref{defmultiperiodgame} is recursive. Since $\widehat s_t+1 > t$, the payoff of $\Game_t$ may depend on the values of $\Game_{t+1}, \ldots, \Game_T$ (which are themselves subgames of $\Game_t$). Intuitively, it is perhaps easier to view the stopping game as a sequence of single period games. If $\Game_t$ is stopped at $t$, then the exercising players receive $X_{k,t}$ while the other players receive
\[
\EP \big(V_{k,t+1}^* \,\big|\, \Filt_t \big) - w_k (\Eset)  \sum_{i \in \Eset} \big( X_{i,t} - \EP \big( V_{i,t+1}^* \,\big|\, \Filt_t\big) \big).
\]
If $\Game_{t}$ is not stopped at time $t$, then it continues on to become the game $\Game_{t+1}$. And finally the game $\Game_T$ always stops at time $T$ as everyone exercises.

\begin{proposition}\label{propstochpayoff}
The expected payoff of $\Game_t$, $\bm{V}_t(s_t)=[V_{1,t}(s_t),\ldots,V_{m,t}(s_t)]$, can be represented using the projection as follows:
\[
\bm{V}_t(s_t)=\EP \big(\proj{\bm{V}_{\widehat s_t+1}^*}{\H_{\Eset(s_t)}}\I_{\{\widehat s_t<T\}} + \bm{X}_T\I_{\{\widehat s_t=T\}} \,\big|\,\Filt_{t}\big),
\]
where $\H_{\Eset(s_t)}$ is the $\Filt_{\widehat s_t}$-measurable hyperplane
\[
\H_{\Eset(s_t)} = \bigg\{ \bm{x} \in \R^m :\,  x_i = X_{i, \widehat s_t},\ \forall i\in \Eset(s_t)\bigg\}.
\]
\end{proposition}
\begin{proof}
This follows immediately from Definition \ref{defmultiperiodgame} and Proposition \ref{thmexproj1}.
\end{proof}

\begin{theorem}\label{thmmultiperiodvalue}
Recursively define the $\Filt_t$-measurable vector $\bm{U}_t=[U_{1,t},\ldots,U_{m,t}]$ by
\begin{align}\label{eq25}
\bm{U}_t=\proj{\EP \big(\bm{U}_{t+1}\,\big|\,\Filt_{t}\big)}{\O_t}, \mbox{ for }0\leq t<T,\quad \bm{U}_T=\bm{X}_T
\end{align}
with $\O_t$ being the $\Filt_t$-measurable orthant
\begin{align}\label{eq26}
\O_t = \bigg\{\bm{x}\in\R^m:\, x_i\geq X_{i,t}, \ 1\leq i \leq m \bigg\},
\end{align}
and the projection $\pi$ is given by Definition \ref{defprojconvention}. 
Define the set of $\FF$-stopping times $\tau$ to be
\begin{align}\label{eq27}
\tau=\big[\tau_1,\ldots,\tau_m\big], \mbox{ where } \tau_i=\inf \big\{ u\in [t,T]: U_{i,u}=X_{i,u} \big\}.
\end{align}
Then
\begin{enumerate}
	\item $\bm{U}_t=\bm{V}_t(\tau)$ taking $\tau$ as a strategy profile of $\Game_t$, and
	\item $\tau$ is an optimal equilibrium of the multi-player stochastic stopping game $\Game_t$, hence $\bm{U}_t=\bm{V}_t^*$ is the value of $\Game_t$.
\end{enumerate}
\end{theorem}
\begin{proof}
The statements are proven simultaneously by backward induction. In the case of $t=T$, $\tau=[T,\ldots,T]$. The game $\Game_T$ is always stopped at time $T$ with the payoff vector $\bm{X}_T=\bm{U}_T=\bm{V}_T(\tau)$ also being the value. Now assume the statements are true for the game $\Game_{t+1}$, so its value is given by 
\begin{align}\label{eq28}
\bm{V}_{t+1}^*=\bm{U}_{t+1}=\bm{V}_{t+1}(\tau).
\end{align}
Note throughout the proof that if the game $\Game_t$ is stopped at time $t$, then it is reduced to a single period stochastic game (cf. Definition \ref{defstochgame}) with payoff vectors $\bm{X}_t$ and $\bm{V}_{t+1}^*$. Denote this single period game by $\Game'$. Also write $\widehat \tau=\tau_1\wedge\cdots \wedge\tau_m$.

\noindent (Statement 1) Case 1: If $\widehat \tau=t$, the game is stopped at time $t$. By Proposition \ref{thmexpectedsoln} and \eqref{eq27}, $\tau$ is an optimal equilibrium of the single period game $\Game'$, whose value is 
\[
\bm{U}_t=\proj{\EP \big(\bm{U}_{t+1}\,\big|\,\Filt_{t}\big)}{\O_t}=\proj{\EP \big(\bm{V}_{t+1}^*\,\big|\,\Filt_{t}\big)}{\O_t}=\proj{\EP \big(\bm{V}_{t+1}^*\,\big|\,\Filt_{t}\big)}{\H_{\Eset(\tau)}}.
\]
The result then follows from Proposition \ref{propstochpayoff}, noting that $\H_{\Eset(\tau)}$ is $\Filt_t$-measurable,
\[
\bm{U}_t=\proj{\EP \big(\bm{V}_{t+1}^*\,\big|\,\Filt_{t}\big)}{\H_{\Eset(\tau)}}=\EP \big(\proj{\bm{V}_{\widehat \tau+1}^*}{\H_{\Eset(\tau)}}\,\big|\,\Filt_{t}\big)=\bm{V}_t(\tau).
\]

Case 2: If $\widehat \tau\geq t+1$, the game is not stopped at time $t$, then by \eqref{eq25}, \eqref{eq26} and \eqref{eq27}, $U_{i,t}>X_{i,t}$ and $\bm{U}_t$ lies in the interior of $\O_t$. Applying the induction hypothesis \eqref{eq28},
\[
\bm{U}_t=\proj{\EP \big(\bm{U}_{t+1}\,\big|\,\Filt_{t}\big)}{\O_t}=\EP \big(\bm{U}_{t+1}\,\big|\,\Filt_{t}\big)=\EP \big(\bm{V}_{t+1}(\tau)\,\big|\,\Filt_{t}\big).
\]
It is sufficient to show $\bm{V}_t(\tau)=\EP \big(\bm{V}_{t+1}(\tau)\,\big|\,\Filt_{t}\big)$, which is true by noting $\widehat \tau\geq t+1$ and applying Proposition \ref{propstochpayoff},
\begin{alignat*}{2}
\bm{V}_t(\tau)&=\EP \big(\EP \big(\proj{\bm{V}_{\widehat \tau+1}^*}{\H_{\Eset(\tau)}}\,\big|\,\Filt_{t+1}\big)\,\big|\,\Filt_{t}\big)=\EP \big(\bm{V}_{t+1}(\tau)\,\big|\,\Filt_{t}\big)&\quad\mbox{if }\widehat \tau &<T,\\
\bm{V}_t(\tau)&=\EP \big(\EP \big(\bm{X}_T\,\big|\,\Filt_{t+1}\big)\,\big|\,\Filt_{t}\big)=\EP \big(\bm{V}_{t+1}(\tau)\,\big|\,\Filt_{t}\big)&\quad\mbox{if } \widehat \tau &=T.
\end{alignat*}

\noindent (Statement 2) By Statement 1, \eqref{eq25} and the induction hypothesis \eqref{eq28}, 
\[
\bm{V}_t(\tau)=\bm{U}_t=\proj{\EP \big(\bm{U}_{t+1}\,\big|\,\Filt_{t}\big)}{\O_t}=\proj{\EP \big(\bm{V}_{t+1}^*\,\big|\,\Filt_{t}\big)}{\O_t}.
\] 
To check $\tau$ is an optimal equilibrium (cf. Definition \ref{optiequilibrium}), we require for each $k\in\Mset$,
\[
V_{k,t}\left(\big[\tau_k, s_{-k}\big]\right)\geq \big[\proj{\EP \big(\bm{V}_{t+1}^*\,\big|\,\Filt_{t}\big)}{\O_t}\big]_k
\geq V_{k,t}\left(\big[s_k, \tau_{-k}\big]\right)
\]
for all $s_k\in\Strat_k$, $s_{-k}\in \Strat_{-k}$. Let $s'=\big[\tau_k, s_{-k}\big], s''=\big[s_k, \tau_{-k}\big]$ be alternative strategy profiles with minimal stopping times $\widehat s', \widehat s''$. 

Case 1: If $\widehat s'=\widehat s''=t$, then both $s'$ and $s''$ can be interpreted as strategy profiles of the single period game $\Game'$. Hence the result follows from Proposition \ref{thmexpectedsoln} because $\big[\proj{\EP \big(\bm{V}_{t+1}^*\,\big|\,\Filt_{t}\big)}{\O_t}\big]_k$ is value of $\Game'$ for player $k$.

Case 2: If $\widehat s'\geq t+1$, then $s'$ is a valid strategy profile of $\Game_{t+1}$. Also we must have $\tau_k \geq t+1$ being a maximin strategy, because it belongs to an optimal equilibrium of $\Game_{t+1}$ by \eqref{eq27} and the induction hypothesis. From Proposition \ref{propstochpayoff},
\begin{alignat*}{2}
\bm{V}_t(s')&=\EP \big(\EP \big(\proj{\bm{V}_{\widehat s'+1}^*}{\H_{\Eset(s')}}\,\big|\,\Filt_{t+1}\big)\,\big|\,\Filt_{t}\big)=\EP \big(\bm{V}_{t+1}(s')\,\big|\,\Filt_{t}\big)&\quad\mbox{if }\widehat s' &<T,\\
\bm{V}_t(s')&=\EP \big(\EP \big(\bm{X}_T\,\big|\,\Filt_{t+1}\big)\,\big|\,\Filt_{t}\big)=\EP \big(\bm{V}_{t+1}(s')\,\big|\,\Filt_{t}\big)&\quad\mbox{if } \widehat s' &=T.
\end{alignat*}
Using the fact that $\tau_k$ is a maximin strategy,
\begin{align}\label{eq29}
V_{k,t}(s')=\EP \big(V_{k,t+1}(s')\,\big|\,\Filt_{t}\big)=\EP \big(V_{k,t}\left(\big[\tau_k, s_{-k}\big]\right)\,\big|\,\Filt_{t}\big)\geq\EP \big(V_{k,t+1}^*\,\big|\,\Filt_{t}\big).
\end{align}
Since $\tau_k\geq t+1$, by \eqref{eq27} we must have 
\begin{align}\label{eq33}
\big[\proj{\EP \big(\bm{V}_{t+1}^*\,\big|\,\Filt_{t}\big)}{\O_t}\big]_k=\big[\proj{\EP \big(\bm{U}_{t+1}\,\big|\,\Filt_{t}\big)}{\O_t}\big]_k>X_{k,t}.
\end{align}
By Proposition \ref{thmexpectedsoln} and \eqref{eq33}, player $k$ does not exercise in the optimal equilibrium of the single period game $\Game'$. Interpreting $\EP \big(V_{k,t+1}^*\,\big|\,\Filt_{t}\big)$ as the expected payoff of player $k$ if no one exercises, and using the definition of optimal equilibrium,
\begin{align}\label{eq30}
\EP \big(V_{k,t+1}^*\,\big|\,\Filt_{t}\big)\geq \big[\proj{\EP \big(\bm{V}_{t+1}^*\,\big|\,\Filt_{t}\big)}{\O_t}\big]_k.
\end{align}
\eqref{eq29} and \eqref{eq30} imply $V_{k,t}(s')\geq \big[\proj{\EP \big(\bm{V}_{t+1}^*\,\big|\,\Filt_{t}\big)}{\O_t}\big]_k$ as required.

Case 3: If $\widehat s''\geq t+1$, by arguments similar to the ones used for \eqref{eq29}
\begin{align}\label{eq31}
V_{k,t}(s'')=\EP \big(V_{k,t+1}(s'')\,\big|\,\Filt_{t}\big)\leq\EP \big(V_{k,t+1}^*\,\big|\,\Filt_{t}\big).
\end{align}
For all $i\neq k$, since $\tau_i\geq t+1$, by \eqref{eq27} we have $\big[\proj{\EP \big(\bm{V}_{t+1}^*\,\big|\,\Filt_{t}\big)}{\O_t}\big]_i=\big[\proj{\EP \big(\bm{U}_{t+1}\,\big|\,\Filt_{t}\big)}{\O_t}\big]_i>X_{i,t}$. By Proposition \ref{thmexpectedsoln}, if $i\neq k$, player $i$ does not exercise in the optimal equilibrium of the single period game $\Game'$. Again interpreting $\EP \big(V_{k,t+1}^*\,\big|\,\Filt_{t}\big)$ as the expected payoff of player $k$ if no one exercises, and using the definition of optimal equilibrium,
\begin{align}\label{eq32}
\EP \big(V_{k,t+1}^*\,\big|\,\Filt_{t}\big)\leq \big[\proj{\EP \big(\bm{V}_{t+1}^*\,\big|\,\Filt_{t}\big)}{\O_t}\big]_k.
\end{align}
\eqref{eq31} and \eqref{eq32} imply $V_{k,t}(s'')\leq \big[\proj{\EP \big(\bm{V}_{t+1}^*\,\big|\,\Filt_{t}\big)}{\O_t}\big]_k$ as required.

Both statements are hence proven and the induction is complete.
\end{proof}

It is possible to further generalise the game by making the weights (hence $a_i$) time dependent and $\FF$-adapted. The weights at time $\widehat s$ will be applied as the game is stopped. Theorem \ref{thmmultiperiodvalue} will analogously hold, with the projection $\pi$ also made time dependent. But for brevity that case will not be included here.

\begin{remark}
The stopping game described by Definition \ref{defmultiperiodgame} is perhaps not the most obvious generalisation of the single period game. A more natural generalisation would be for the non-exercising player $k$ to receive
\[
X_{k,T} - w_k (\Eset(s_t))  \sum_{i \in \Eset(s_t)} \big( X_{i,\widehat s_t} - X_{k,T}\big),\quad \widehat s_t<T,
\]
as the game is stopped. That is, using $X_{k,T}$ instead of the value $V^*_{k,\widehat s_t+1}$ of $\Game_{t+1}$. But even in deterministic cases, this does not always produce optimal equilibria in pure strategies. For example, consider a game with $\bm{X}_0=[-1,-1,0], \bm{X}_1=[-2,-2,4], \bm{X}_2=[0,0,0]$ and $a_1=a_2=a_3=1/3$. Player 3 will always want to exercise at time 1, while there is a prisoner's dilemma between players 1 and 2 at time 0. This game has two Nash equilibria (with different payoffs) but no optimal equilibria in pure strategies.
\end{remark}

\subsection{Quitting Games}\label{secquitgames}

A quitting game is an alternative formulation of the multiple period case. As opposed to a stopping game, a quitting game does not end when one player stops or exercises. Instead, the non-exercising players continue the game and may exercise at a later date. We focus on the deterministic case here, as the stochastic case does not always produce optimal equilibria (see Remark \ref{remstochquit}).

\begin{definition}\label{defquittinggame}
A deterministic \emph{multi-player stochastic quitting game} $\Game$, with players indexed by $\Mset=\{1,2,\ldots,m\}$, is specified by the following inputs:
\begin{itemize}
	\item The vectors $\bm{X}_t=[X_{1,t}, \ldots, X_{m,t}]$, $t=0,\ldots,T$;
	\item The weights $w_k(\Eset)=\frac{a_k}{1-\sum_{i\in\Eset}a_i}$, where $a_i>0$ and $\sum_{i\in\Mset} a_i \leq 1$.
\end{itemize}
The rules of the game are as follows. If player $k$ exercises at time $t$ in the interval $[0,T-1]$ (or $s_k=t$), he receives a payoff of $X_{k,t}$. If player $k$ does not exercise before time $T$ (or $s_k=T$), he receives $X_{k,T}-w_k (\Eset(s)) \sum_{i \in \Eset(s)} (X_{i,t}-X_{i,T})$ at time $T$, where $\Eset(s)$ is the set of players exercising before time $T$ and $s=[s_1,\ldots,s_m]$. In other words, the payoff vector $\bm{V}(s)=[V_1(s),\ldots,V_m(s)]$ is given by
\[
V_k(s)=
\begin{cases}
 X_{k,s_k}, & s_k<T,\\
 X_{k,T} - w_k(\Eset(s)) \sum_{i \in \Eset(s)} (X_{i,s_i}-X_{i,T}), & s_k=T.
\end{cases}
\]

\end{definition}

\begin{theorem}\label{thmquitsoln}
The quitting game $\Game$ has a unique value given by
\[
\bm{V}^*=\proj{\bm{X}_t}{\O}
\]
where $\O$ is the orthant defined by
\[
\O = \bigg\{\bm{x}\in\R^m:\, x_i\geq \max_{0\leq t\leq T-1} X_{i,t}, \ 1\leq i \leq m \bigg\}
\]
and the projection $\pi$ is given by Definition \ref{defprojconvention}. A possible optimal equilibrium $s^*=[s_1^*,\ldots,s_m^*]$ is given by
\[
s_i^*=\min\big\{t\in[0,T]: \proj{\bm{X}_t}{\O}_i\leq X_{i,t} \big\}.
\]
\end{theorem}

\begin{proof}
Consider a single period game $\Game'$ with $X_k'=\max_{0\leq t\leq T-1} X_{k,t}$ and $P_k'=X_{k,T}$. By Theorem \ref{thmsoln3}, $\bm{V}^*$ is the value of $\Game'$. If player $k$ exercises then $\proj{\bm{X}_t}{\O}=X_k'=\max_{0\leq t\leq T-1} X_{k,t}$. If player $k$ does not exercise then $V_k^*=X_{k,T}-w_k(\Eset(s^*))D(s^*)\leq X_{k,T}$ (recall $D(s^*)\geq 0$ from Lemma \ref{lemposdiff}). Either way $s_k^*$ is well-defined, and the exercise decision corresponds to the optimal equilibrium of $\Game'$. Hence it is easy to check that $\bm{V}^*=\bm{V}(s^*)$.

Now to see $s^*$ is an optimal equilibrium of the quitting game $\Game$, as per Definition \ref{optiequilibrium}, first we check that it is a Nash equilibrium, or
\[
V_i\left(\big[s_k^*, s_{-k}^*\big]\right)\geq V_k\left(\big[s_k, s_{-k}^*\big]\right), \quad \forall \, s_k\in\Strat_{-k}.
\]
If $s_k^*<T$ and player $k$ exercises. He cannot improve his payoff by exercising at another time since $X_{k,s_k^*}$ is maximal, neither can he improve by not exercising since $s^*$ corresponds to a Nash equilibrium in $\Game'$.

If instead $s_k^*=T$ and player $k$ does not exercise, it is sufficient to check that he cannot improve by exercising at any time, or $\bm{V}^*\geq \max_{0\leq t\leq T-1} X_{k,t}=X_k'$. This is certainly true since, once again, $s^*$ corresponds to a Nash equilibrium in $\Game'$. Therefore $s^*$ is a Nash equilibrium of $\Game$.

To complete the proof, we are left to check
\[
V_k\left(\big[s_k^*, s_{-k}\big]\right)\geq V_k\left(\big[s_k^*, s_{-k}^*\big]\right), \quad \forall \, s_{-k}\in\Strat_{-k}.
\]
If $s_k^*<T$ and player $k$ exercises, his payoff is fixed and cannot be decreased by the action of other players. If instead $s_k^*=T$ and player $k$ does not exercise, write $s=\big[s_k^*, s_{-k}\big]$, then
\begin{align}\label{eq35}
V_k(s)&=X_{k,T}-w_k(\Eset(s))\sum_{i \in \Eset(s)} (X_{i,s_i}-X_{i,T})\nonumber\\
&\geq X_{k,T}-w_k(\Eset(s))\sum_{i \in \Eset(s)} \left(\max_{0\leq t\leq T-1}X_{i,t}-X_{i,T}\right)\nonumber\\
&=P_k'-w_k(\Eset(s))\sum_{i \in \Eset(s)} (X_i'-P_i') = V_k(s')
\end{align}
where $s'$ is the strategy profile in which any exercising player under $s$ chooses to exercise for maximal payoff $X_i'$ instead. But $s'$ corresponds to a strategy profile in $\Game'$ with player $k$ not exercising. Since $s^*$ is an optimal equilibrium of $\Game'$, we have $V_k(s')\geq V_k(s^*)$. Combining with \eqref{eq35}, it implies $V_k(s)\geq V_k(s^*)$, as required.
\end{proof}

One thing to note about Theorem \ref{thmquitsoln} is that it doesn't specify the amount of information available to the players regarding the exercise decisions of others. Unlike a stopping game, the strategies in a quitting game can also depend on the observable actions of other players. However, if we denote the total information available to player $k$ over time by the filtration $\FF_k=\{\Filt_{k,t}, t=0,\ldots, T\}$, then the strategy $s_k$ is an $\FF_k$ stopping time. Theorem \ref{thmquitsoln} shows that, in the quitting game, regardless of how much or little any player observes about the actions of others, the value of the game is fixed. An optimal equilibrium attaining the value can be chosen independently.

\begin{remark}
If $\Game$ is a \emph{perfect information} quitting game, that is, $s_i$ is an $\FF_j$ stopping time for any $i\neq j$, then a \emph{subgame perfect optimal equilibrium} can be constructed. Denote the set of exercising player up to time $t-1$ by $\Eset_{t-1}(s)$ and the set of remaining players by $\Mset_t=\Mset\setminus\Eset_{t-1}(s)$. By Lemma \ref{lemsubgame}, the quitting game $\Game_t$ on the interval $[t,T]$ is a subgame amongst the remaining players $\Mset_t$, but with following variable modifications:
\begin{itemize}
\item $X_{k,u}'=X_{k,u}$, for $t\leq u\leq T-1$, $k\in\Mset_t$;
\item $X_{k,T}'=X_{k,T}-w_k(\Eset_{t-1}(s))\sum_{i\in\Eset_{t-1}(s)}(X_{i,s_i}-X_{i,T})$, for $k\in\Mset_t$;
\item Weights defined by $a_k'=w_k(\Eset_{t-1}(s))=\frac{a_k}{1-\sum_{i\in\Eset_{t-1}(s)}a_i}$, for $k\in\Mset_t$; and
\item Strategy profile $s_t=[s_{k,t},\ k\in\Mset_t]\in \Strat_t$.
\end{itemize}

A subgame perfect optimal equilibrium $s^{**}$ is an optimal equilibrium for any reachable subgame of $\Game$. In particular it is also an optimal equilibrium and attains the value $\bm{V}(s^{**})=\bm{V}^*$. The optimal equilibrium $s^*$ constructed in Theorem \ref{thmquitsoln} is not necessarily subgame perfect, as the strategy $s_k^*$ of player $k$ does not take the actions of others into account. For example, assume another player $i$ deviates from $s_i^*$ by exercising too early. The strategy $s_k^*$ does not adjust to punish the mistake. Even though the value of $\bm{V}_k^*$ is still guaranteed, player $k$ misses the chance to guarantee an even higher payoff, created by the sub-optimal deviation of player $i$.

The subgame perfect optimal equilibrium can be constructed recursively, by backward induction with respect to both time and remaining set of players. Specifically, for the quitting game subgame $\Game_t$, the subgame perfect optimal equilibrium $s_t^{**}=\big[s_{k,t}^{**},\ k\in\Mset_t\big]$ is given by $s_{k,T}^{**}=T$ and
\[
s_{k,t}^{**}=t\I_{\{\proj{X_{k,T}}{\O_t}_k=X_{k,t}\}}+s_{k,t+1}^{**}\I_{\{\proj{X_{k,T}}{\O_t}_k>X_{k,t}\}},\quad t<T
\]
where $\O_t=\big\{x_i\geq \max_{t\leq u\leq T-1}X_{i,u},\ i\in\Mset_t \big\}$ is an $\R^{|\Mset_t|}$ orthant, the projection $\pi$ is defined by Definition \ref{defprojconvention} with modifications for $\Mset_t$, and $s_{k,t+1}^*$ is the player $k$ subgame perfect optimal equilibrium strategy of the subgame $\Game_{t+1}$. Finally $s^{**}=s_{0}^{**}$ is the subgame perfect optimal equilibrium of the perfect information quitting game $\Game$.

On the other hand, in an \emph{imperfect information} quitting game, where the players' actions are partially or completely hidden to others, it is not always possible to determine the current subgame $\Game_t$. Therefore subgame perfect optimal equilibria may not exist. But by using the same idea as above, each player can construct a strategy which is the optimal equilibrium strategy in all observable subgames. And as mentioned, the lack of subgame perfection does not change the value of the quitting game.
\end{remark}

\begin{remark}\label{remstochquit}
In the stochastic case, unfortunately equilibria may not exist in pure strategies. For example, a game with $a_1=a_2=a_3=1/3$, $\Filt_1=\Filt_2=\{\emptyset,\{\omega_1\},\{\omega_2\},\Omega\}$, $\mathbb{P}(\omega_1)=\mathbb{P}(\omega_2)=1/2$, and the payoffs given by $\bm{X}_0=[2.1,3.5,-50]$, $\bm{X}_1(\omega_1)=[-50,-50,-5.05]$, $\bm{X}_1(\omega_2)=[4,-50,-50]$ and $\bm{X}_2(\omega_1)=\bm{X}_2(\omega_2)=[0,5,-5]$.
\end{remark}

\appendix
\section{Appendix}\label{secappendix}

\subsection{Proof of Propositions \ref{thmweightswuc} and \ref{propalwaysnash}}\label{secappendix1}

Before proving the Proposition \ref{thmweightswuc}, we require the following lemmas.
\begin{lemma}\label{lemweightswuc0}
Given a subset $\Eset\subset\Mset$ with $0\leq |\Eset(s)| \leq m-2$ and $i, j \notin \Eset(s)$, let $\Eset'=\Eset\cup\{j\}$. If $s$ and $s'$ are strategy profiles with $\Eset(s)=\Eset$ and $\Eset(s')=\Eset'$ (hence player $j$ exercises in $s'$ but not $s$), Then
\begin{gather}\label{eqweightswuc}
(V_i(s)-V_i(s'))+w_i(\Eset')(V_j(s)-V_j(s'))=\left\{w_i(\Eset')[1-w_j(\Eset)]-w_i(\Eset)\right\}D(s).
\end{gather}
\end{lemma}
\begin{proof}
This follows directly from the definition of $\bm{V}$, noting that $D(s')=D(s)+(X_j-P_j)$. Start with
\begin{align*}
V_i(s)-V_i(s')&=w_i(\Eset')D(s')-w_i(\Eset)D(s)\\
&=(w_i(\Eset')-w_i(\Eset))D(s)+w_i(\Eset')(X_j-P_j),\\
w_i(\Eset')(V_j(s)-V_j(s')) &= w_i(\Eset')(P_j-w_j(\Eset)D(s) - X_j)\\
&=-w_i(\Eset')w_j(\Eset)D(s)-w_i(\Eset')(X_j-P_j).
\end{align*}
Adding the expressions yields the desired result.
\end{proof}

\begin{lemma}\label{lemweightswuc}
The game $\Game$ is WUC for all choices $\bm{X}$ and $\bm{P}$, if and only if both of the following Conditions hold:
\begin{enumerate}
	\item For any $\Eset\subset\Mset, 1\leq |\Eset| \leq m-2$ and $i, j \notin \Eset$,
\[
w_i(\Eset\cup\{j\})(1-w_j(\Eset))=w_i(\Eset);
\]
	\item For any $\Eset'\subset\Mset, 1\leq |\Eset'| \leq m-1$ and $i \notin \Eset'$,
\[
w_i(\Eset')>0.
\]
\end{enumerate}
\end{lemma}
\begin{proof} We prove the statement in three parts.

\noindent(WUC $\Rightarrow$ Condition 1) Take $s$ and $s'$ to be strategy profiles with $\Eset(s)=\Eset$ and $\Eset(s')=\Eset'=\Eset\cup\{j\}$. If the game $\Game$ is WUC for all $\bm{X}$ and $\bm{P}$, then
\[
V_j(s)=V_j(s') \Longrightarrow V_i(s)=V_i(s').
\]
By Lemma \ref{lemweightswuc0}, \eqref{eqweightswuc} becomes
\[
\left\{w_i(\Eset')[1-w_j(\Eset)]-w_i(\Eset)\right\}D(s) = 0.
\]
When $|\Eset|\geq 1$, we can choose $\bm{X}, \bm{P}$ so that $D(s)\neq 0$. Hence
\[
w_i(\Eset')[1-w_j(\Eset)] = w_i(\Eset)
\]
and Condition 1 is proven.

\noindent(Condition 1 $+$ WUC $\Rightarrow$ Condition 1 $+$ Condition 2) Now assume Condition 1 holds, hence $w_i(\Eset')[1-w_j(\Eset)] - w_i(\Eset)=0$ for $|\Eset|\geq 1$. Note that if $|\Eset|= 0$, then $D(s)=0$. In either case, \eqref{eqweightswuc} always simplifies to
\[
(V_i(s)-V_i(s'))=w_i(\Eset')(V_j(s')-V_j(s)).
\]
But the $\Game$ being WUC requires $V_j(s')-V_j(s)>0 \Longrightarrow V_i(s)-V_i(s')\geq 0$. Since the weights are required to be non-zero, we have $w_i(\Eset')>0$ for all $1\leq |\Eset'| \leq m-1$ and Condition 2 is proven.

\noindent(WUC $\Leftarrow$ Condition 1 $+$ Condition 2) As before, Condition 1 quickly gives 
\[
(V_i(s)-V_i(s'))=w_i(\Eset')(V_j(s')-V_j(s)).
\]
When Condition 2 also holds, it's easily checked that the game is indeed WUC.
\end{proof}

\begin{proof}[Proposition \ref{thmweightswuc}]
It is sufficient to completely solve the system presented in Conditions 1 and 2 of Lemma \ref{lemweightswuc}. Consider the case where $m\geq 4$. 
By Condition 2, we have $0<w_i(\{j\})<1$ for all $i\neq j$. By Condition 1,
\[
\frac{w_i(\{j\})}{1-w_k(\{j\})}=w_i(\{j,k\})=\frac{w_i(\{k\})}{1-w_j(\{k\})} \Longleftrightarrow \frac{w_i(\{j\})}{w_i(\{k\})}=\frac{1-w_k(\{j\})}{1-w_j(\{k\})}. 
\]
Since the only the left hand side depend on $i$, we have, for $i,j,k,l$ distinct,
\[
\frac{w_i(\{j\})}{w_i(\{k\})}=\frac{w_l(\{j\})}{w_l(\{k\})}. 
\]
This system of equations has the parametric solution $w_i(\{j\})=a_ib_j$, with $a_i, b_i \neq 0$. Substituting back,
\[
\frac{a_ib_j}{1-a_kb_j}=\frac{a_ib_k}{1-a_jb_k} \Longleftrightarrow a_j+\frac{1}{b_j}=a_k+\frac{1}{b_k} \Longleftrightarrow a_i+\frac{1}{b_j}=c
\]
where $c$ is a constant for all $i$. Solving for $b_j$ yields 
$
w_i(\{j\})=a_ib_j=\frac{a_i}{c-a_j}.
$
Scale all $a_i$ by a factor of $1/c$,
\[
w_i(\{j\})=\frac{a_i}{1-a_j}.
\]
Substituting into Condition 1 while recursively incrementing the size of $\Eset$, we obtain
\[
w_k(\Eset)=\frac{a_k}{1-\sum_{i\in\Eset}a_i}.
\]
Condition 2 adds the restrictions $a_k>0, \sum_{i\neq k} a_i<1$. This solution can easily be checked to always satisfy both Conditions 1 and 2.

The cases for $m=2, 3$ can be easily solved to obtain the same solutions.
\end{proof}

\begin{proof}[Proposition \ref{propalwaysnash}]
First note that for $i\neq j$
\[
w_i(\Mset\setminus\{i\})=\frac{a_i}{1-\sum_{j\neq i} a_j}>1.
\]
On the other hand, as before, $w_i(\Eset)<1$ if $|\Eset|\leq m-2$.

We will simply check all the possibilities. If $\{3,\ldots,m\}\subseteq \Eset$ or if there are only two players, then the cases are the following.
\begin{itemize}
\item If both players 1 and 2 exercise, then $V_1=V_2=0$, but player 2 can receive $-1+w_2(\Mset\setminus\{2\})>0$ if he doesn't exercise.
\item If only player 1 exercises, then $V_1=0$, but player 1 can receive $P_1=1$ if he doesn't exercise.
\item If only player 2 exercises, then $V_1=1-w_1(\Mset\setminus\{1\})<0$, but player 1 can receive $X_1=0$ if he also exercises.
\item If neither player 1 or 2 exercises, then $V_2=-1$, but player 2 can receive $X_2=0$ if he also exercises. 
\end{itemize}

If some player $k\neq 1,2$ doesn't exercise, so $k\notin \Eset$, then the cases are the following.
\begin{itemize}
\item If both players 1 and 2 exercise, then $V_1=V_2=0$, but $|\Eset\setminus\{1\}|\leq m-2$ and player 1 can receive $1-w_1(\Eset\setminus\{1\})>0$ if he doesn't exercise.
\item If only player 1 exercises, then $|\Eset|\leq m-2$ and $V_2=-1+w_1(\Eset)<0$, but player 2 can receive $X_2=0$ if he exercises.
\item If only player 2 exercises, then $V_k=0-w_k(\Eset)<0$, but player $k$ can receive $X_k=0$ if he also exercises.
\item If neither player 1 or 2 exercises, then $V_2=-1$, but player 2 can receive $X_2=0$ if he also exercises. 
\end{itemize}

In all cases, the pure strategy profile is not an equilibrium.
\end{proof}

\subsection{Proof of Lemmas \ref{lemsubgame} and \ref{lemposdiff}}\label{secappendix3}
\begin{proof}[Proof of Lemma \ref{lemsubgame}]
Let $s'\in \Strat_{\Mset'}$ be any strategy profile of the subgame $\Game_{\Mset'}$ and $\Eset'=\Eset'(s')$ be the corresponding set of exercising players. Let $s\in\Strat$ be a matching strategy profile of the original game $\Game$, so
\[
s=[s_k=s_k', k\in\Mset'; s_k=0, k\notin\Mset']
\]
and $\Eset(s)=\Eset\cup\Eset'$. It is sufficient to check that for any $k\in\Mset'$, the payoff $V'_k(s')$ of the subgame matches the payoff $V_k(s)$ of the original game.

First note that the weights of $\Game_{\Mset'}$ can be written as
\begin{gather}\label{eq6}
w_k'(\Eset')=\frac{a_k'}{1-\sum_{i\in\Eset'}a_i'}=\frac{w_k(\Eset)}{1-\sum_{i\in\Eset'}w_i(\Eset)}\\
\label{eq7}=\frac{\frac{a_k}{1-\sum_{i\in\Eset}a_i}}{1-\sum_{i\in\Eset'}\frac{a_i}{1-\sum_{j\in\Eset}a_j}}=\frac{a_k}{1-\sum_{i\in\Eset\cup\Eset'}a_i}=w_k(\Eset(s)).
\end{gather}

If $k\in\Eset'$, then $V'_k(s')=X_k'=X_k=V_k(s)$. Otherwise,
\begin{align}\label{eq8}
V'_k(s') &= P_k' - w_k'(\Eset') \sum_{i\in\Eset'}(X_i'-P'_i)\nonumber\\
&=P_k-w_k(\Eset)\sum_{i\in\Eset}(X_i-P_i) - w_k'(\Eset') \sum_{i\in\Eset'}\left(X_i-P_i+w_i(\Eset) \sum_{j\in\Eset}(X_j-P_j)\right)\nonumber\\
&=P_k-\left(w_k(\Eset)+w_k'(\Eset')\sum_{i\in\Eset'}w_i(\Eset)\right)\sum_{i\in\Eset}(X_i-P_i) - w_k'(\Eset') \sum_{i\in\Eset'}(X_i-P_i).
\end{align}
Rearranging \eqref{eq6} as $w_k(\Eset)+w_k'(\Eset')\sum_{i\in\Eset'}w_i(\Eset)=w_k'(\Eset')$, we can rewrite \eqref{eq8} as
\begin{align*}
V'_k(s')&=P_k-w_k'(\Eset')\sum_{i\in\Eset}(X_i-P_i) - w_k'(\Eset') \sum_{i\in\Eset'}(X_i-P_i)\\
&=P_k-w_k'(\Eset')\sum_{i\in\Eset(s)}(X_i-P_i)\\
&=P_k-w_k(\Eset(s))\sum_{i\in\Eset(s)}(X_i-P_i)\qquad\qquad\qquad\text{by \eqref{eq7}}\\
&=V_k(s),
\end{align*}
as required.
\end{proof}

\begin{proof}[Lemma \ref{lemposdiff}]
Assume the contrary, so $D(s^*)<0$. Then there must exists a player $k \in \Eset(s^*)$ with $X_{k}-P_k<0$.
Now if player $k$ chooses not to exercise, his payoff will be
$
V_k(s') =  P_k - w_k(\Eset(s'))  \sum_{i\in  \Eset(s')} (X_{i}-P_i),
$
where $s'$ is the modified strategy profile. Note that $s' = [1, s_{-k}^*]$ and $s^*= [0, s_{-k}^*]$
and thus $\Eset(s')= \Eset(s^*) \setminus \{k\}$.
Since $s^*$ is an equilibrium, we have $V_k(s') \leq V_k(s^*) = X_k$. Therefore,
\begin{align}\label{eq9}
X_k - V_k(s') = X_k - P_k + w_k(\Eset(s')) \sum_{i\in  \Eset(s')} (X_{i}-P_i)  \geq 0 .
\end{align}
Since $X_{k}-P_k<0$, \eqref{eq9} implies $\sum_{i\in \Eset(s')} (X_{i}-P_i) \geq 0$. Recall in Definition \ref{assumpw3}, $0 < w_k(\Eset(s')) \leq 1$. Therefore
\[
D(s^*)= \sum_{i\in \Eset(s^*)} (X_{i} - P_i) \geq X_k - P_k + w_k(\Eset(s')) \sum_{i\in  \Eset(s')} (X_{i}-P_i) \geq 0 ,
\]
contradicting the assumption of $D(s^*)<0$.
\end{proof}

\subsection{Proof of Theorem \ref{thmsoln3}}\label{secappendix2}
The following lemma is a standard result of projection in linear algebra. It is used throughout the other proofs.
\begin{lemma}\label{lemlinalg}
In $\R^m$, if $\K$ is a hyperplane, the projection $\pi$ is orthogonal, that is, $\proj{\bm{P}}{\K}$ is the unique vector in $\K$ such that
\[
\left\langle\proj{\bm{P}}{\K}-\bm{P}, \bm{Q}-\proj{\bm{P}}{\K} \right\rangle=0,\quad \forall\, \bm{Q}\in\K.
\]
Furthermore, if $\J$ is a convex subset of the hyperplane $\K$, then
\[
\proj{\bm{P}}{\J}=\proj{\proj{\bm{P}}{\K}}{\J}.
\]
\end{lemma}

\begin{proof}[Proof of Proposition \ref{thmexproj1}]
The vector $\bm{V} (s)$
\[
\bm{V} (s) = \big[ V_i (s) = X_{i},\, i\in \Eset(s),\, V_i (s) = P_i - w_i(\Eset(s))D (s),\, i\notin\Eset(s)\big].
\]
certainly lies in the hyperplane
\[
\H_{\Eset(s)} = \bigg\{ \bm{x} \in \R^m :\,  x_i = X_{i}\ \mbox{for every}\ i\in \Eset (s) \bigg\}.
\]
So it is sufficient to check that
\begin{align*}
\bm{v} := \bm{P}-\bm{V}(s)&=\big[ v_i=P_i-X_{i},\, i\in\Eset(s),\, v_i=w_i(\Eset(s)) D (s),\, i\notin\Eset(s)\big]
\end{align*}
is orthogonal to $\H_{\Eset(s)}$. Let
$
\bm{u} := \big[ u_i=0,\, i\in\Eset(s)\big]
$
be any vector lying entirely in $\H_{\Eset(s)}$. Then $\langle \bm{u}, \bm{v} \rangle$ evaluates to
\begin{align*}
&\hphantom{={}} \sum_{i\notin\Eset(s)} \frac{u_iw_i(\Eset(s)) D (s)}{a_i} + \frac{\left(\sum_{i\notin\Eset(s)} u_i \right)\left(\sum_{i\in \Eset(s)} \left(P_i-X_{i}\right) + \sum_{i\notin\Eset(s)}w_i(\Eset(s)) D (s)\right)}{1-\sum_{i=1}^m a_i}\\
&= \frac{\sum_{i\notin\Eset(s)} u_i D (s)}{1-\sum_{i\in\Eset(s)}a_i} + \frac{\left(\sum_{i\notin\Eset(s)} u_i \right)\left(-D(s) + \frac{\sum_{i\notin\Eset(s)} a_i}{1-\sum_{i\in\Eset(s)}a_i} D (s)\right)}{1-\sum_{i=1}^m a_i}\\
&= \frac{\sum_{i\notin\Eset(s)} u_i D (s)}{1-\sum_{i\in\Eset(s)}a_i}\left(1 + \frac{-\left(1-\sum_{i\in\Eset(s)}a_i\right)+\sum_{i\notin\Eset(s)} a_i}{1-\sum_{i=1}^m a_i}\right)\\
&= \frac{\sum_{i\notin\Eset(s)} u_i D (s)}{1-\sum_{i\in\Eset(s)}a_i}\left(1 -1 \right)=0
\end{align*}
as required.
\end{proof}

Before proving Theorem \ref{thmsoln3}, a couple more lemmas are needed.

\begin{lemma}
\label{LemSimplexProj0}
Assume that $\bm{P}\in \O$. Then $\proj{\bm{P}}{\H_\Aset}\in \O$ for any subset $\Aset\subseteq\Mset$.
\end{lemma}

\begin{proof}
By Theorem \ref{thmexproj1}, the projection $\proj{\bm{P}}{\H_\Aset}$ corresponds to the payoff vector when
$\Aset$ is the set of exercising players. Let $s$ be the corresponding strategy profile. In particular, for any $i\in\Aset$
\[
\left[\proj{\bm{P}}{\H_\Aset}\right]_i = X_{i} \leq P_i .
\]
and thus $D(s) = \sum_{i \in \Aset } (X_i-P_i) \leq 0$. Consequently, for any $j \in \Mset \setminus \Aset $
\[
\left[\proj{\bm{P}}{\H_\Aset}\right]_j = P_j - w_j(\Eset(s)) (s) D(s) \geq P_j \geq X_{j},
\]
and thus $\proj{\bm{P}}{\H_\Aset}\in \O$.
\end{proof}

\begin{lemma}
\label{LemSimplexProj1}
Let $k \in \Mset $. If $\proj{\bm{P}}{\O}\notin\H_{\{k\}}$ then $P_k>X_{k}$. Equivalently, if $P_k \leq X_{k}$ then $\proj{\bm{P}}{\O}\in\H_{\{k\}}$.
\end{lemma}

\begin{proof}
Suppose that $P_k \leq X_{k}$ and assume that
$\proj{\bm{P}}{\O}\notin\H_{\{k\}}$. Then the projection $\bm{Q}=\proj{\proj{\bm{P}}{\O}}{\H_{\{k\}}}$ is still in $\O$  (by Lemma \ref{LemSimplexProj0})
and it is distinct from $\proj{\bm{P}}{\O}$ (since $\proj{\bm{P}}{\O}\notin\H_{\{k\}}$). We will show that
\begin{align} \label{distq}
\norm{\bm{P}-\bm{Q}} < \norm{\bm{P}-\proj{\bm{P}}{\O}} ,
\end{align}
which contradicts the definition of $\proj{\bm{P}}{\O}$.

\begin{figure}[ht]
\centering
\begin{tikzpicture}[line cap=round,line join=round,>=triangle 45,x=1.2cm,y=1.2cm]
\clip(-3,-0.5) rectangle (6,3.5);
\draw [domain=-2.5:4.5] plot(\x,0);
\draw (2,3.46)-- (0,0);
\draw (-2,0)-- (0.77,1.33);
\draw (0.77,1.33)-- (0.77,0);
\draw[color=black] (-2,-0.3) node {$\bm{P}$};
\draw[color=black] (0.3,1.58) node {$\proj{\bm{P}}{\O}$};
\draw[color=black] (0.77,-0.3) node {$\bm{Q}$};
\draw[color=black] (5,-0.05) node {$\H_{\{k\}}$};
\end{tikzpicture}
\caption{\label{fig1}$P_k = X_k$}
\end{figure}
In the case of $P_k = X_k$, as shown in Figure \ref{fig1}, we have $\bm{P},\bm{Q}\in\H_{\{k\}}$ and $\proj{\bm{P}}{\O}-\bm{Q}$ being orthogonal to $\bm{P}-\bm{Q}$. Hence
\[
\norm{\bm{P}-\bm{Q}}^2 < \norm{\bm{P}-\bm{Q}}^2+\norm{\proj{\bm{P}}{\O}-\bm{Q}}^2 = \norm{\bm{P}-\proj{\bm{P}}{\O}}^2.
\]

\begin{figure}[ht]
\centering
\begin{tikzpicture}[line cap=round,line join=round,>=triangle 45,x=1.2cm,y=1.2cm]
\clip(-1,-1.5) rectangle (6,3.5);
\draw [domain=-1:5] plot(\x,0);
\draw (2,3.46)-- (0,0);
\draw (2,-1)-- (0.77,1.33);
\draw (2,-1)-- (0.77,0);
\draw [domain=-1:5] plot(\x,-1);
\draw (0.77,1.33)-- (0.77,-1);
\draw[color=black] (2,-1.3) node {$\bm{P}$};
\draw[color=black] (0.3,1.58) node {$\proj{\bm{P}}{\O}$};
\draw[color=black] (0.5,-0.3) node {$\bm{Q}$};
\draw[color=black] (0.77,-1.3) node {$\bm{R}$};
\draw[color=black] (5.5,-0.05) node {$\H_{\{k\}}$};
\draw[color=black] (5.5,-1.01) node {$\widehat \H_{\{k\}}$};
\end{tikzpicture}
\caption{\label{fig2}$P_k < X_k$}
\end{figure}
To establish \eqref{distq} in the case $P_k < X_k$, as shown in Figure \ref{fig2}, we introduce a hyperplane $\widehat \H_{\{k\}}$ parallel to $\H_{\{k\}}$ by setting
\[
\widehat \H_{\{k\}} = \bigg\{ \bm{x} \in \R^m :\,  x_k =P_k \bigg\},
\]
so that, in particular, $\bm{P} \in \widehat \H_{\{k\}}$. Let $\bm{R}=\proj{\proj{\bm{P}}{\O}}{\widehat \H_{\{k\}}}$,
so that also
\[
\bm{R}=\pi_{\widehat \H_{\{k\}}} \big( \pi_{\H_{\{k\}}} ( \pi_{\O } (\bm{P})) \big)= \proj{\bm{Q}}{\widehat \H_{\{k\}}}.
\]
Since  $P_k < X_{k}$,  $\bm{R} \in \widehat \H_{\{k\}}$ and $\pi_{\O } (\bm{P}) \in \O \setminus \H_{\{k\}}$ lie on opposite sides of the
hyperplane $\H_{\{k\}}$.  It is thus clear that
\begin{align} \label{distq1}
\norm{\bm{R}-\bm{Q}} <
\norm{\bm{R}-\bm{Q}} + \norm{\bm{Q}-\proj{\bm{P}}{\O}} = \norm{\bm{R}-\proj{\bm{P}}{\O}}.
\end{align}

Finally, since $\bm{P}-\bm{R}$ is orthogonal to both $\bm{R}-\bm{Q}$ and $\bm{R}-\proj{\bm{P}}{\O}$, we have
\[
\norm{\bm{P}-\bm{Q}}^2 = \norm{\bm{P}-\bm{R}}^2 + \norm{\bm{R}-\bm{Q}}^2
\]
and
\[
\norm{\bm{P}-\proj{\bm{P}}{\O}}^2  = \norm{\bm{P}-\bm{R}}^2 + \norm{\bm{R}-\proj{\bm{P}}{\O}}^2.
\]
Therefore, \eqref{distq1} implies \eqref{distq}, as required.
\end{proof}

\begin{proof}[Theorem \ref{thmsoln3}]
Begin by noting that in the subgame $\Game_{\Mset'}$ (where $\Mset'=\Mset\setminus\Eset$) defined in Lemma \ref{lemsubgame}, the variables $P_k'$ can be rewritten as
\[
P_k'=P_k-w_k(\Eset)\sum_{i\in\Eset}(X_i-P_i)=\big[\proj{\bm P}{\H_\Eset}\big]_k
\]
according to Proposition \ref{thmexproj1}. The mapping $\phi:\H_\Eset \rightarrow \R^{m-|\Eset|}$, defined by discarding the coordinates with indices in $\Eset$, is an isometry to the space of $\Game_{\Mset'}$ payoffs. Endowed it with the norm
\[
\norm{\bm{x}}'= \left(\sum_{i\in\Mset'} \left( \frac{x_i^2}{a_i'}\right)+ \frac{\left(\sum_{i\in\Mset'} x_i \right)^2}{1-\sum_{i\in\Mset'} a_i'}\right)^{\frac{1}{2}}.
\]
and let $\pi'$ be the corresponding projection function.

Back to the main proof, it is sufficient to show that the strategy profile $s^*$ defined by
\[
s_i^*=0\quad\Longleftrightarrow\quad\big[\proj{\bm{P}}{\O}\big]_i=X_i
\]
is a Nash equilibrium (hence an optimal equilibrium since $\Game$ is WUC). This is done using the same induction from the proof of Theorem \ref{thmsoln}, but with a few additions. The base case of $m=1$ can be easily checked.

Consider $m\geq 2$. If $P_i > X_i$ for all $i$, then $\bm{P}$ lies in the interior of $\O$. So $\proj{\bm{P}}{\O}=\bm{P}$ and $s^*=\bm{1}$ is an equilibrium. If $P_k \leq X_k$ for some $k$, consider the $m-1$ player subgame $\Game_{\{-k\}}$. By the induction hypothesis, $s'\in\Strat_{-k}$ defined by 
\[
s_i'=0\quad\Longleftrightarrow\quad\big[\pi'_{\O'}{\bm{P}'}\big]_i=X_i,\quad \forall\,i\in\Mset\setminus\{k\}
\]
is an equilibrium of $\Game_{\{-k\}}$. Apply the isometry $\phi^{-1}$, then Lemma \ref{lemlinalg},
\[
\big[\pi'_{\O'}\left(\bm{P}'\right)\big]_i=\big[\proj{\proj{\bm P}{\H_{\{k\}}}}{\O\cap\H_{\{k\}}}\big]_i=\big[\proj{\bm P}{\O\cap\H_{\{k\}}}\big]_i,\quad \forall\,i\in\Mset\setminus\{k\} .
\]

By Lemma \ref{LemSimplexProj1}, $P_k \leq X_k$ implies $\proj{\bm{P}}{\O}\in\H_{\{k\}}$. Hence $\proj{\bm P}{\O\cap\H_{\{k\}}}=\proj{\bm P}{\O}$ and $s'$ can be rewritten as
\[
s_i'=0\quad\Longleftrightarrow\quad\big[\proj{\bm{P}}{\O}\big]_i=X_i,\quad \forall\,i\in\Mset\setminus\{k\}.
\]
Finally $\big[\proj{\bm{P}}{\O}\big]_k=X_k$ implies $s_k^*=0$, therefore
\[
s^*=\big[s_k^*=0,\ s_{-k}^*=s'].
\]
By the proof of Theorem \ref{thmsoln}, $s^*$ must be an equilibrium of $\Game$, as required.
\end{proof}

\bibliographystyle{apalike}
\bibliography{references}
\end{document}